\documentclass[journal,letterpaper,twoside,twocolumn]{IEEEtran}
\usepackage{amsmath,amssymb,graphicx,psfrag,cite}
\usepackage[normalem]{ulem} 
\usepackage[usenames,dvipsnames]{color} 
\usepackage{balance}
\usepackage{mathrsfs}   
\usepackage{paralist}
\usepackage{bbm}

\graphicspath{{figure/}}

\newcommand{\argmin}{\mathop{\rm arg~min}\limits}

\usepackage{color}

\newcommand{\rev}[1]{{\color{blue}#1}}

  \renewcommand{\sout}[1]{}  \renewcommand{\rev}[1]{#1}

\title{Performance Monitoring for Live Systems \\with Soft FEC and Multilevel Modulation}
\author{Tsuyoshi~Yoshida,~\IEEEmembership{Member,~IEEE,} Mikael~Mazur,~\IEEEmembership{Member,~IEEE,} Jochen~Schr\"oder,~\IEEEmembership{Member,~IEEE,} Magnus~Karlsson,~\IEEEmembership{Fellow,~OSA;~Senior~Member,~IEEE,} and Erik~Agrell,~\IEEEmembership{Fellow,~IEEE} \thanks{Manuscript received November 15, 2019\rev{; revised February 4, 2020}. This work was partly presented at ECOC 2019 \cite{yoshida_2019_ecoc}.}%
	\thanks{This work was funded in part by ``Massive Parallel and Sliced Optical Network (MAPLE),'' the Commissioned Research of National Institute of Information and Communications Technology (NICT), Japan (project number 20401), and the Swedish Research Council, VR (grant 2014-06138).}%
	\thanks{T.~Yoshida is with Information Technology R\&D Center, Mitsubishi Electric Corporation, Kamakura, 247-8501 Japan (e-mails:  Yoshida.Tsuyoshi@ah.MitsubishiElectric.co.jp). He also belongs to Osaka University, Suita, Osaka, 505-0871 Japan.}%
	\thanks{M.~Mazur, J.~Schr\"oder, M.~Karlsson, and E.~Agrell are with the Fiber-Optic Communications Research Center (FORCE), Chalmers University of Technology, SE-41296 Gothenburg, Sweden.}
	\thanks{Copyright (c) 2020 IEEE. Personal use of this material is permitted. However, permission to use this material for any other purposes must be obtained from the IEEE by sending a request to pubs-permissions@ieee.org.}
}%
\begin{document}
\IEEEspecialpapernotice{(Top-Scored)}
\maketitle

\begin{abstract}
Performance monitoring is an essential function for margin measurements in live systems. Historically, system budgets have been described by the Q-factor converted from the bit error rate (BER) under binary modulation and direct detection. The introduction of hard forward error correction (FEC) did not change this. In recent years, technologies have changed significantly to comprise coherent detection, multilevel modulation and soft FEC. In such advanced systems, different metrics such as (nomalized) generalized mutual information (GMI/NGMI) and asymmetric information (ASI) are regarded as being more reliable. On the other hand, Q budgets are still useful because pre-FEC BER monitoring is established in industry for live system monitoring. 

The pre-FEC BER is easily estimated from available information of the number of flipped bits in the FEC decoding, which does not require knowledge of the transmitted bits that are unknown in live systems. Therefore, the use of metrics like GMI/NGMI/ASI for performance monitoring has not been possible in live systems. However, in this work we propose a blind soft-performance estimation method. Based on a histogram of log-likelihood-values without the knowledge of the transmitted bits, we show how the ASI can be estimated. 

We examine the proposed method experimentally for $16$- and $64$-ary quadrature amplitude modulation (QAM) and probabilistically shaped $16$-, $64$-, and $256$-QAM in recirculating loop experiments. We see a relative error of $\rev{3.6}\%$, which corresponds to around $0.5\,\mathrm{dB}$ signal-to-noise ratio difference for binary modulation, in the regime where the ASI is larger than the assumed FEC threshold. For this proposed method, the digital signal processing circuitry requires only the minimal additional function of storing the L-value histograms before the soft FEC decoder.
\end{abstract}

\begin{IEEEkeywords}
Bit error rate, bitwise decoding, forward error correction, modulation, mutual information, optical fiber communication, performance monitoring, probabilistic shaping.
\end{IEEEkeywords}

\section{Introduction}
\label{sec:intro}
Optical fiber communications are expected to carry traffic at high data rates with high reliability. Thus a very low bit error rate (BER) is required, such as $10^{-9}$ or $10^{-12}$ in early systems and currently $10^{-15}$ \cite{ITU-T_G.975.1}. Following a standard \cite{ITU-T_G.977}, system vendors have used margins from the threshold Q-factor converted from the corresponding BER to cope with potential extra impairments due to fiber nonlinearity, polarization/wavelength-dependent phenomena, system aging and variation, component imperfections, etc. In recent years, there has been a trend to reduce margin allocation for efficient networking in cases of dynamic network reconfiguration \cite{oda_2017}. Then, real-time performance monitoring, whose trends are well summarized in \cite{dong_2016}, is becoming more important than ever. 
When examining the performance with known test bits, one can easily quantify the performance at the receiver. However, once the system starts to carry live traffic, the true transmitted bits are unknown at the receiver. Then, the performance must be monitored blindly, without knowledge of the transmitted data. Optical spectrum analysis is one of the key instruments that can establish the system status. More recently, optical modulation analyzers have became essential in optical transceiver production. However, to install such instruments for entire systems is unrealistic due to cost reasons. Thus there have been many studies of performance monitoring with minimal or low-cost hardware since the days of on--off keying modulated (OOK) systems. The main monitoring target can be for example optical signal-to-noise ratio (OSNR) or chromatic and polarization-mode dispersion \cite{kilper_2004}. Neural networks have been proposed for such monitoring and also applied to detect imperfect modulation in the transmitter \cite{wu_2009}. Monitoring with coherent detection was proposed for direct-detection systems \cite{fu_2005}, and after the deployment of coherent communications with digital signal processing (DSP), the digital coherent receiver is used for performance monitoring \cite{hauske_2009}. In-band OSNR monitoring \cite{adams_2006,zhao_2014} is also important, partly because it is inaccessible to conventional optical spectrum analyzers. Further, in dense wavelength division multiplexing with channel spacing close to Nyquist limit, the optical background noise can be difficult to estimate\cite{bosco_2011}. Another important topic is to distinguish the noise from amplified spontaneous emission (ASE) and nonlinear noise, such as broadband four-wave mixing under nonlinear fiber-optic transmission \cite{sinkin_2013}, which has been studied extensively\cite{xiang_2018,kashi_2018}. 
Very recently, data-analysis-based power profile estimation over multiple spans was demonstrated \cite{tani_2019}.

Among the various targets of performance monitoring, the BER is one of the most basic and essential ones in network operation.
For systems with OOK without forward error correction (FEC) with a few $\mathrm{dB}$s of system margin, the BER is usually too low to be measurable in a reasonable time. 
Thus the BER or the corresponding Q-factor at the optimum decision threshold was estimated from the BERs using nonideal decision thresholds, also called ``bath-tub curves'' \cite{bergano_1993}. Alternatively, the Q-factor could be estimated from the statistics (mean and standard deviations) of the eye-diagram. 
Since the deployment of systems with hard FEC \cite{song_2002}, the performance has been quantified by the pre-FEC BER (or Q-factor) and compared with the boundary value required for quasi-error-free transmission after FEC decoding, the so-called \emph{FEC limit}. A pre-FEC BER is then estimated by the bit flipping ratio in the FEC decoder. 

In the last decade, coherent detection, DSP \cite{roberts_2009_jlt}, soft FEC \cite{chang_2010_cm} and multilevel modulation have been widely deployed, as summarized in \cite{alvarado_2018_jlt}. More recently, probabilistic shaping (PS) with reverse concatenation \cite{bocherer_2015,buchali_2016} have gained much attention for its capacity-approaching performance. Then more reliable performance metrics to predict post-FEC performance must be considered. At the same time, information-theoretic metrics such as mutual information (MI) \cite{schmalen_2017}, generalized MI (GMI) \cite{alvarado_2015}, normalized GMI (NGMI) \cite{cho_2017}, achievable binary code (ABC) rate \cite{bocherer_2017_arxiv}, and asymmetric information (ASI) \cite{yoshida_2017_ptl}\rev{, achievable FEC rate \cite{bocherer_2019_jlt}} have been introduced. To compute these metrics, the transmitted bits must be known, and their application to performance monitoring in live systems without this knowledge is not straightforward. 
At typical operation points, which are near or beyond the ``waterfall region'' of modern soft FEC codes, the SNR can be relatively low, e.g., $10\,\mathrm{dB}$, which can make the performance estimate unreliable. It is thus valuable to verify and quantify in simulations and experiments the accuracy of these modern performance metrics in a context with powerful codes operating at relatively low SNR levels.

This work is an invited extension of \cite{yoshida_2019_ecoc}. We propose a performance-monitoring technique applicable to systems carrying live traffic with soft FEC and multilevel modulation such as high-order quadrature amplitude modulation (QAM) under bitwise decoding. We estimate the ASI, which is known as a good predictor of post-FEC BER with soft FEC, multilevel modulation, and PS \cite{yoshida_2017_ptl,zhang_2019}. The proposed estimation method is validated experimentally using various combinations of modulation formats and shaping codes. The main extensions from \cite{yoshida_2019_ecoc} lie in a more detailed theoretical background, \rev{additional examinations under different FEC thresholds and launch powers,} Q-factor estimation based on slightly modified and improved statistics, 
and additional back-to-back experimental results to investigate the cause of the estimation errors.

This paper is organized as follows: Sec.~\ref{sec:pre} describes the theoretical background required to understand the  proposed method, which  is then presented in Sec.~\ref{sec:prp}. Sec.~\ref{sec:exp} describes the experiment and the results, and Sec.~\ref{sec:cnc} concludes the work.

\section{Preliminaries}
\label{sec:pre}
\begin{figure}[tb]
	\begin{center}
		\setlength{\unitlength}{.6mm} %
		\scriptsize
		\includegraphics[scale=0.48]{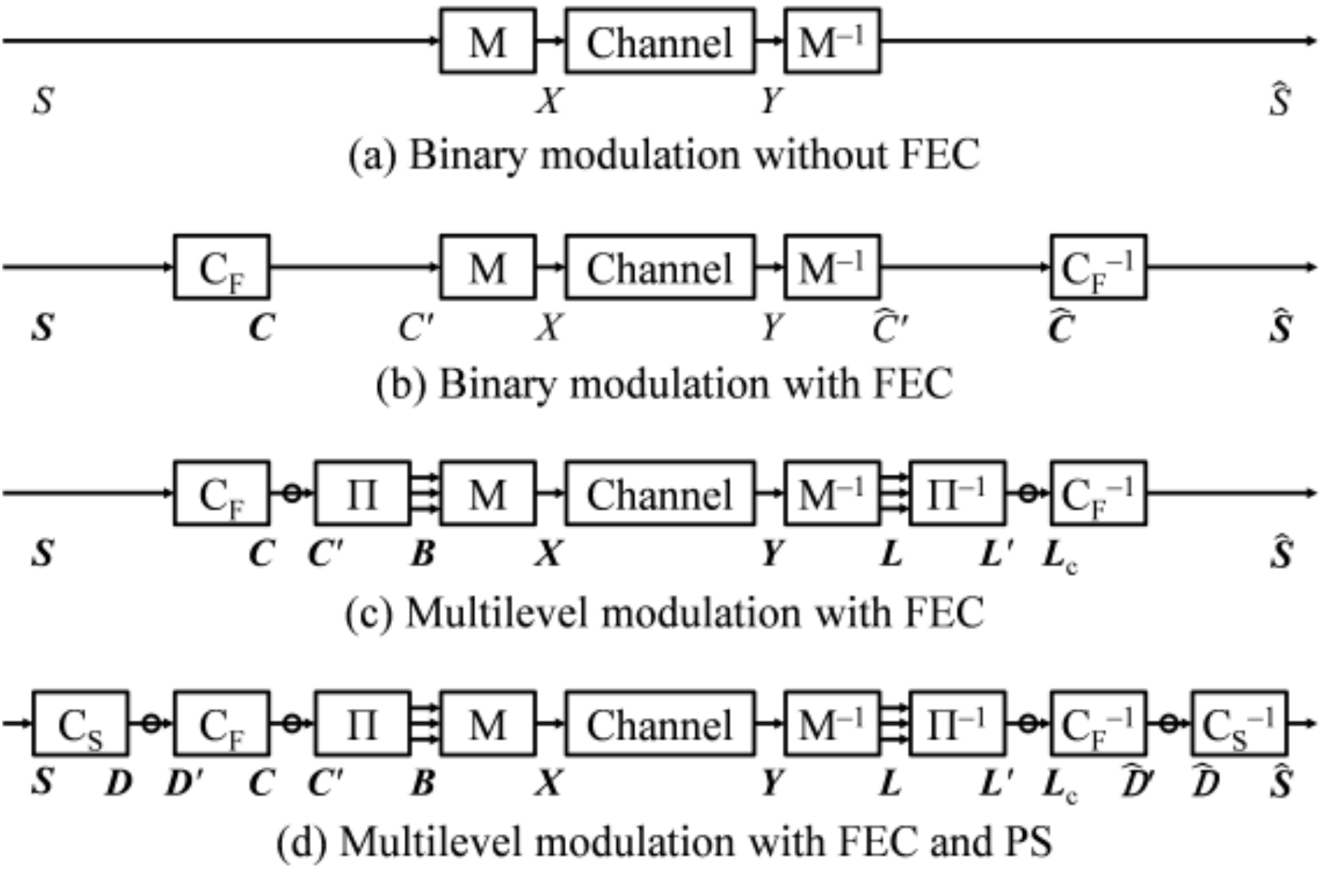} \\
		\vspace{-0.3cm}		
		\caption{System models: (a) binary modulation without FEC, (b) binary modulation with FEC, (c) multilevel modulation with FEC, and (d) multilevel modulation with FEC and PS. The functions M, $\mathrm{C}_{\mathrm{F}}$, $\Pi$, and $\mathrm{C}_{\mathrm{S}}$ are symbol mapping, FEC encoding, bit-interleaving, and PS encoding at the transmitter. The functions at the receiver are their inverse operations. }
		\label{fig:sys}
	\end{center}
	\vspace{-0.3cm}
\end{figure}

In this section, we summarize system models and relevant performance metrics for, first, legacy systems with OOK, direct detection and hard FEC, and second, more modern systems with coherent detection and possibly high-order QAM, PS and soft FEC. Fig.~\ref{fig:sys} shows some typical deployable system models. The models and performance metrics for binary modulation are explained in Sec.~\ref{subsec:prebin}, and those for multilevel modulation in Sec.~\ref{subsec:premlt}, respectively.

\subsection{Performance in systems with binary modulation}
\label{subsec:prebin}
Optical fiber communications have supported binary modulation for a long time, such as OOK, differential binary/quaternary phase-shift keying (BPSK/QPSK), and coherent BPSK/QPSK with polarization-division multiplexing.
Fig.~\ref{fig:sys}(a) shows the system model for binary modulation without FEC, which was employed for OOK systems. The source bits $S\in\mathbb{B}$ are mapped into binary symbols $X\in\mathbb{R}$, and received symbols $Y\in\mathbb{R}$ are demapped into estimated bits $\hat{S}\in\mathbb{B}$, where $\mathbb{B}$ and $\mathbb{R}$ denote the binary number set and the real number set, resp\rev{ectively}.
Such early systems were not equipped with FEC, but the raw BER had to be quasi-error-free, which in practice meant lower than $10^{-9}$ or $10^{-12}$, while $10^{-15}$ is the target in the most recent standards. To quantify such a low BER easier or more intuitively, the BER was converted into the (SNR-like) Q-factor
\begin{IEEEeqnarray}{rCL}
	\label{eq:Qber}
	Q_{\mathrm{BER}} & \triangleq & \sqrt{2} \mathrm{erfc}^{-1} ( 2\cdot \mathrm{BER} ) .
\end{IEEEeqnarray}
The Q-factor $Q_{\mathrm{BER}}$ shows more linear relationship with (the electrical) signal-to-noise ratio (SNR) and OSNR than the BER.  
For uniform binary modulation, $Q_{\mathrm{BER}}^2$ gives the SNR required for a given BER over the Gaussian channel. 
Historically, the Q-factor was related to the eye-opening of binary modulation. From the mean $\mu_b$ (denoting the average location of the upper and lower ``rails'' of the eye diagram) and standard deviation $\sigma_b$ (representing the width of said rails) of the sampled signals for the transmitted bits $b\in\mathbb{B}$, the ``statistical'' Q-factor was defined as
\begin{IEEEeqnarray}{rCL}
	\label{eq:preQst}
	Q_{\mathrm{st}} &\triangleq& \frac{|\mu_0 - \mu_1|}{\sigma_0+\sigma_1}.
\end{IEEEeqnarray}
In the case of uniform binary modulation over the Gaussian channel, $Q_{\mathrm{st}} = Q_{\mathrm{BER}}$. \rev{While this holds for a thermal-noise-limited channel, a direct-detection optical channel limited by ASE noise is non-Gaussian, with a noncentral chi-square distribution, making $Q_{\mathrm{st}} \neq Q_{\mathrm{BER}}$.}\footnote{\sout{While this holds for a thermal-noise-limited channel, a direct-detection optical channel limited by ASE noise is  non-Gaussian, with a noncentral Chi-square distribution, making $Q_{\mathrm{st}} \neq Q_{\mathrm{BER}}$.}\rev{In coherent BPSK/QPSK systems, the Gaussian channel model is applicable even with optical amplifiers.}} There are many performance monitoring methods for OOK from such statistics. Without the knowledge of the transmitted bits, $\mu_b$ and $\sigma_b$ are estimated as $\hat{\mu}_b$ and $\hat{\sigma}_b$, based on an assumed suitable decision criterion to estimate the transmitted bit $b$ for each sample. Then the statistical Q-factor can be estimated as
\begin{IEEEeqnarray}{rCL}
	\label{eq:preQst_pd}
	\hat{Q}_{\mathrm{st}} \triangleq \frac{|\hat{\mu}_0 - \hat{\mu}_1|}{\hat{\sigma}_0+\hat{\sigma}_1} \approx Q_{\mathrm{st}},
\end{IEEEeqnarray}
which could be obtained directly from the eye diagram \sout{of the received data}\rev{measured by an oscilloscope}, and hence became a popular system quality metric due to its simplicity.
When the BER is less than $10^{-9}$, the estimated statistics $\hat{\mu}_b$ and $\hat{\sigma}_b$, and the resulting estimated Q-factor $\hat{Q}_{\mathrm{st}}$ are reliable. 

With increased throughput, FEC was introduced for efficiency and reliability\cite{song_2002,chang_2010_cm}. Fig.~\ref{fig:sys}(b) shows the system model in that case. At the transmitter, source bits $\boldsymbol{S}\in\mathbb{B}^{k}$ are encoded by the FEC into $\boldsymbol{C}\in\mathbb{B}^n$, where $n$ and $k$ now denote the FEC codeword length and the number of information bits, resp\rev{ectively}. The encoded bits $C' \in\mathbb{B}$ are then mapped into binary symbols $X\in\mathbb{R}$. At the receiver, the received symbols $Y\in\mathbb{R}$ are demapped into the estimated encoded bits $\hat{C}'\in\mathbb{B}$.\footnote{In most cases, hard FEC has been employed in systems with binary modulation, including 100 Gb/s coherent detection systems, while long-reach coherent BPSK/QPSK systems employed soft FEC\cite{chang_2010_cm}. Then, the received symbols $Y$ are demapped into L-values as shown in Sec.~\ref{subsec:premlt}.} The estimated encoded bits for a FEC codeword $\hat{\boldsymbol{C}}\in\mathbb{B}^n$ are decoded into the estimated source bits $\hat{\boldsymbol{S}}\in\mathbb{B}^k$. 
In this system, the relevant performance metric is now the BER after FEC decoding\footnote{If PS is employed, the BER after PS decoding, if that is placed after the FEC decoding, is more relevant \cite{yoshida_ecoc_2018,yoshida_jlt_2019,amari_2019_arxiv}.}, or post-FEC BER, 
\begin{IEEEeqnarray}{rCL}
	\label{eq:postBER}
	\mathit{BER}_{\mathrm{post}} &\triangleq& \sum_{b\in \mathbb{B}} P_{S,\hat{S}} (b,1-b),
\end{IEEEeqnarray}
where ($S,\hat{S})$ denotes a random pair of bits in ($\boldsymbol{S},\hat{\boldsymbol{S}})$
and $P_{S,\hat{S}} (b,1-b)$ denotes the joint probability of $S = b$ and $\hat{S} = 1-b$. Recent system requirements on $\mathit{BER}_{\mathrm{post}}$ can be as low as $10^{-15}$.
However, to measure such a low $\mathit{BER}_{\mathrm{post}}$ in real time is difficult and time-consuming. Instead, the
BER before FEC decoding, the pre-FEC BER, 
\begin{IEEEeqnarray}{rCL}
	\label{eq:preBER}
	\mathit{BER}_{\mathrm{pre}} &\triangleq& \sum_{b\in \mathbb{B}} P_{C,\hat{C}}(b,1-b) ,
\end{IEEEeqnarray}
where ($C,\hat{C})$ denotes a random pair of bits in ($\boldsymbol{C},\hat{\boldsymbol{C}})$,   
is often measured and converted into a Q-factor by \eqref{eq:Qber}. Since $\mathit{BER}_{\mathrm{pre}}$ is usually higher in systems with FEC than in systems without FEC, estimating the Q-factor via \eqref{eq:preQst_pd} is less reliable. 

A threshold value of $\mathit{BER}_{\mathrm{pre}}$ into the FEC decoder for error-free operation is also characterized in the FEC design and called the \emph{Q limit}. In the standardized Q budget \cite{ITU-T_G.977}, 
the starting point is $Q_{\mathrm{BER}}$ at a given received OSNR at the beginning-of-life of the system, and the penalties are then quantified relative to this Q-factor, never to reach the Q limit before the system's end-of-life.

\subsection{Performance in systems with soft binary FEC and multilevel modulation}
\label{subsec:premlt}

Fig.~\ref{fig:sys}(c) shows the system model for multilevel modulation with soft
binary FEC, i.e., bit-interleaved coded modulation
(BICM)\cite{zehavi_1992_tcom,caire_1998_tit,fabregas_2008,bicmbook}. This system
model is relevant for 400\,Gb/s standards of 400ZR \cite{400zr} and openROADM
\cite{openROADM}, e.g., 400 Gb/s 16-QAM with concatenated (soft) Hamming and
(hard) staircase codes \cite{lyub_2018}.
This BICM system can be further generalized to include PS in Fig.~\ref{fig:sys}(d), so in the following we describe Fig.~\ref{fig:sys}(d).
Source bits $\boldsymbol{S}\in\mathbb{B}^{k_{\mathrm{ps}}}$ are encoded into $\boldsymbol{D}\in\mathbb{B}^{n_{\mathrm{ps}}}$ by PS encoding. After reframing, the PS-encoded bits $\boldsymbol{D}'\in\mathbb{B}^{k}$ are encoded by FEC into the bits $\boldsymbol{C}\in\mathbb{B}^n$. The FEC-encoded bits $\boldsymbol{C}'\in\mathbb{B}^{m}$ are interleaved into bits $\boldsymbol{B}=(B_1,\ldots,B_m)\in\mathbb{B}^{m}$, where $m$ denotes the number of bits per complex symbol. The bits $\boldsymbol{B}$ are mapped into symbols $\boldsymbol{X}\in\mathbb{R}^2$.

In the receiver, the received symbols $\boldsymbol{Y}$ are demapped to L-values $\boldsymbol{L} = (L_1,\ldots,L_m)\in\mathbb{L}^m$ as \cite[Eqs.~(3.31)--(3.32), (3.39)]{bicmbook}
\begin{IEEEeqnarray}{rCL}
	\label{eq:Lvalue}
	L_i(\boldsymbol{y}) \triangleq \left[ \ln \frac{q_{B_i,\boldsymbol{Y}}(0,\boldsymbol{y})}{q_{B_i,\boldsymbol{Y}}(1,\boldsymbol{y})}\right]_\mathbb{L} ,
\end{IEEEeqnarray}
where $[l]_\mathbb{L}$ denotes quantization of a real number $l$ to the nearest value in $\mathbb{L}$, and $q_{B_i,\boldsymbol{Y}}(b,\boldsymbol{y})$ is the joint probability distribution assumed in the demapper. It can be factorized as $q_{B_i,\boldsymbol{Y}}(b,\boldsymbol{y}) = P_{B_i}(b)q_{\boldsymbol{Y}\mid B_i}(\boldsymbol{y} \!\mid\!  b)$, where $P_{B_i}(b)$ denotes the probability of bit $B_i$ being $b$ and $q_{\boldsymbol{Y}\mid B_i}(\boldsymbol{y} \!\mid\!  b)$ is the auxiliary channel assumed in the demapper. 
Furthermore, the demapping function quantizes both the input symbols $\boldsymbol{Y}$ (to typically 64 to 1024 levels in each dimension)
and output L-values $\boldsymbol{L}$, 
using uniform mid-rise quantization. Specifically, the discrete set of L-values
is $\mathbb{L} \triangleq \{-l_{\mathrm{max}},-l_{\mathrm{max}}+\Delta
l,-l_{\mathrm{max}}+2\Delta l,\ldots, l_{\mathrm{max}}\}$, where
$l_{\mathrm{max}} \triangleq (n_{\mathrm{bin}}-1)\Delta l/2$, $n_{\mathrm{bin}}$ is the number of
bins, and $\Delta l$ is the step size.

After deinterleaving $\boldsymbol{L}$ into $\boldsymbol{L}'\in\mathbb{L}^m$ and framing, the L-values per FEC codeword $\boldsymbol{L}_{\mathrm{c}}\in\mathbb{L}^n$ are decoded by the FEC decoder into $\boldsymbol{D}\in\mathbb{B}^k$. The decoded bits per PS codeword $\hat{\boldsymbol{D}}{}'\in\mathbb{B}^{n_{\mathrm{ps}}}$ are decoded by the PS decoder into the estimated source bits $\hat{\boldsymbol{S}}\in\mathbb{B}^{k_{\mathrm{ps}}}$. 
If there is no PS coding, $k_{\mathrm{ps}}=k$, then the source bits $\boldsymbol{S}$ are fed directly into the FEC encoder, and the FEC decoder output gives the estimated source bits $\hat{\boldsymbol{S}}$. 

In analogy with \eqref{eq:postBER}, the post-FEC BER for systems with BICM and PS is $\mathit{BER}_{\mathrm{post}} = \sum_{b\in \mathbb{B}} P_{D',\hat{D}'} (b,1-b)$, where ($D',\hat{D}')$ denotes a random pair of bits in ($\boldsymbol{D}',\hat{\boldsymbol{D}}{}')$.
As predictors of the post-FEC BER under BICM with PS, \rev{more} useful metrics \rev{than pre-FEC BER (or {$Q_{\text{BER}}$})} are NGMI\cite{cho_2017,cho_2019_jlt}, ABC rate\cite{bocherer_2017_arxiv}, \sout{and} ASI\cite{yoshida_2017_ptl}\rev{, and achievable FEC rate \cite{bocherer_2019_jlt}}. \rev{The post-FEC BER curves were compared in \cite{cho_2017,yoshida_2017_ptl,yoshida_2019_arxiv,yoshida_2020_ofc}.} They are all equivalent under matched decoding but not under mismatched decoding\cite{yoshida_2019_arxiv}. The difference between NGMI and ASI was first observed in \cite{zhang_2019}. 
The definitions of GMI \rev{\cite[Eq.~(20)]{kaplan_1993_aeu},} \cite[Eq.~(21)]{alvarado_2015}, NGMI \cite[Eq.~(6)]{cho_2017}, \cite[Eq.~(14)]{cho_2019_jlt}, and ASI \rev{$I_{\mathrm{a}}$} \cite[Eq.~(11)]{yoshida_2017_ptl} are
\begin{IEEEeqnarray}{rCL}
	\label{eq:gmi}
	\mathit{GMI}&\triangleq& \max_{s>0} I_{\rev{q,}s}^{\mathrm{gmi}} (\boldsymbol{B};\boldsymbol{Y})  , \\
	\label{eq:Isgmi}
	\!\!\!\! I_{\rev{q,}s}^{\mathrm{gmi}} (\boldsymbol{B};\boldsymbol{Y}) &\triangleq& \mathbb{E}_{\boldsymbol{B},\boldsymbol{Y}} \! \left[ \log_2 \frac{ \rev{q_{\boldsymbol{Y} \mid \boldsymbol{B}}(\boldsymbol{y} \! \mid \! \boldsymbol{b})}^\rev{s}}{\sum_{\boldsymbol{b\in\mathbb{B}^m}} P_{\boldsymbol{B}}(\boldsymbol{b}) \rev{q_{\boldsymbol{Y} \mid \boldsymbol{B}}(\boldsymbol{y} \! \mid \! \boldsymbol{b})}^{\rev{s}} } \right] , \\
	\label{eq:NGMI}
	\mathit{NGMI} &\triangleq& 1- \frac{1}{m} \left( \mathbb{H}(\boldsymbol{B}) - \mathit{GMI} \right) , \\
	\label{eq:ASI}
	\rev{I_{\mathrm{a}}} &\triangleq& 1 - \mathbb{H}( L_{\mathrm{a}} \mid | L_{\mathrm{a}} | ) , 
\end{IEEEeqnarray}
where $\mathbb{H}(\cdot)$ and $\mathbb{E}[\cdot]$ denote entropy and expectation. $L_{\mathrm{a}}$ is the symmetrized \emph{a posteriori} L-values $L_{\mathrm{a}} \triangleq (-1)^b L$, where $L=L_I$ for a random bit tributary $I$, selected uniformly from $\{1,\ldots,m\}$, and $b$ is the corresponding transmitted bit in $B_I$ \cite[Eq.~(33)]{alvarado_2015}, \cite[Eq.~(10)]{yoshida_2017_ptl}.\footnote{In \cite{alvarado_2015,yoshida_2017_ptl}, the L-values were defined over the set of real numbers, whereas in this paper, we define them over the discrete set $\mathbb{L}$. Therefore, entropy is used in \eqref{eq:ASI} instead of differential entropy as in \cite{yoshida_2017_ptl}.}
\sout{Under the bit metric decoding} \rev{In a bitwise receiver, which} we assume in this paper, the \sout{symbol decoding metric $q(\boldsymbol{b},\boldsymbol{y})$} \rev{channel assumed in the demapper $q_{\boldsymbol{Y}\mid\boldsymbol{B}}(\boldsymbol{y}\!\mid\!\boldsymbol{b})$} is \cite[Eq.~(22)]{buchali_2016} 
\begin{IEEEeqnarray}{rCL}
	\label{eq:qs}
	\rev{q_{\boldsymbol{Y}\mid\boldsymbol{B}}(\boldsymbol{y}\!\mid\!\boldsymbol{b})} &=& \frac{1}{P_{\boldsymbol{B}}(\boldsymbol{b})}\prod_{i=1}^m q_{B_i,\boldsymbol{Y}}(b_{\rev{i}},\boldsymbol{y}) \rev{,}
\end{IEEEeqnarray}
\rev{where $\boldsymbol{b}=(b_1 b_2 \ldots b_m)$. The scaling parameter $s$ controls the auxiliary channel, and the optimum $s$ \rev{($s_{\mathrm{o}}$)} gives the GMI as $I_{\rev{q},s_{\rev{\mathrm{o}}}}(\boldsymbol{B};\boldsymbol{Y})$.}
\sout{As shown in [41], } 
\sout{the NGMI and the ASI can rewritten as}
\sout{From (12)} 
\sout{and (13), }
\sout{it is clear that
$\mathit{ASI} = \mathit{NGMI}$}
\rev{The NGMI and the ASI are equivalent} under matched decoding ($s_{\rev{\mathrm{o}}}=1$), but not necessarily so under mismatched decoding ($s_{\rev{\mathrm{o}}} \neq 1$) \rev{\cite{yoshida_2017_ptl}}. 
Furthermore, it can be shown that the NGMI is equivalent to the ABC rate under matched decoding \cite{yoshida_2019_arxiv}.
\sout{Reversing (13), }
\sout{an achievable information rate (AIR) with BICM with PS can be computed from the ASI as }
\sout{When considering the rate loss in PS coding [21, Sec.~V-B], [42, Eq.~(4)],} 
\sout{, we should subtract it from the } \sout{r.h.s.} \sout{ of (14).} 
As an example of a practical implementation, the SNR of the auxiliary channel is set to a fixed FEC threshold value, as in \cite{yoshida_2016}. Then there may be a gap of up to several $\mathrm{dB}$s between the SNRs of the received symbols carrying live traffic and that of the auxiliary channel \cite{yoshida_2019_arxiv}.
\rev{The ASI is computed from such mismatched and quantized L-values, and it characterizes the performance just before the FEC decoder in deployable systems.}

\section{Proposed method}
\label{sec:prp}

In this section we explain the configuration of the receiver side DSP and the operation principle of the proposed blind ASI monitoring method. 

\subsection{Configuration}

\begin{figure}[tb]
	\begin{center}
		\setlength{\unitlength}{.6mm} %
		\scriptsize
		\includegraphics[scale=0.35]{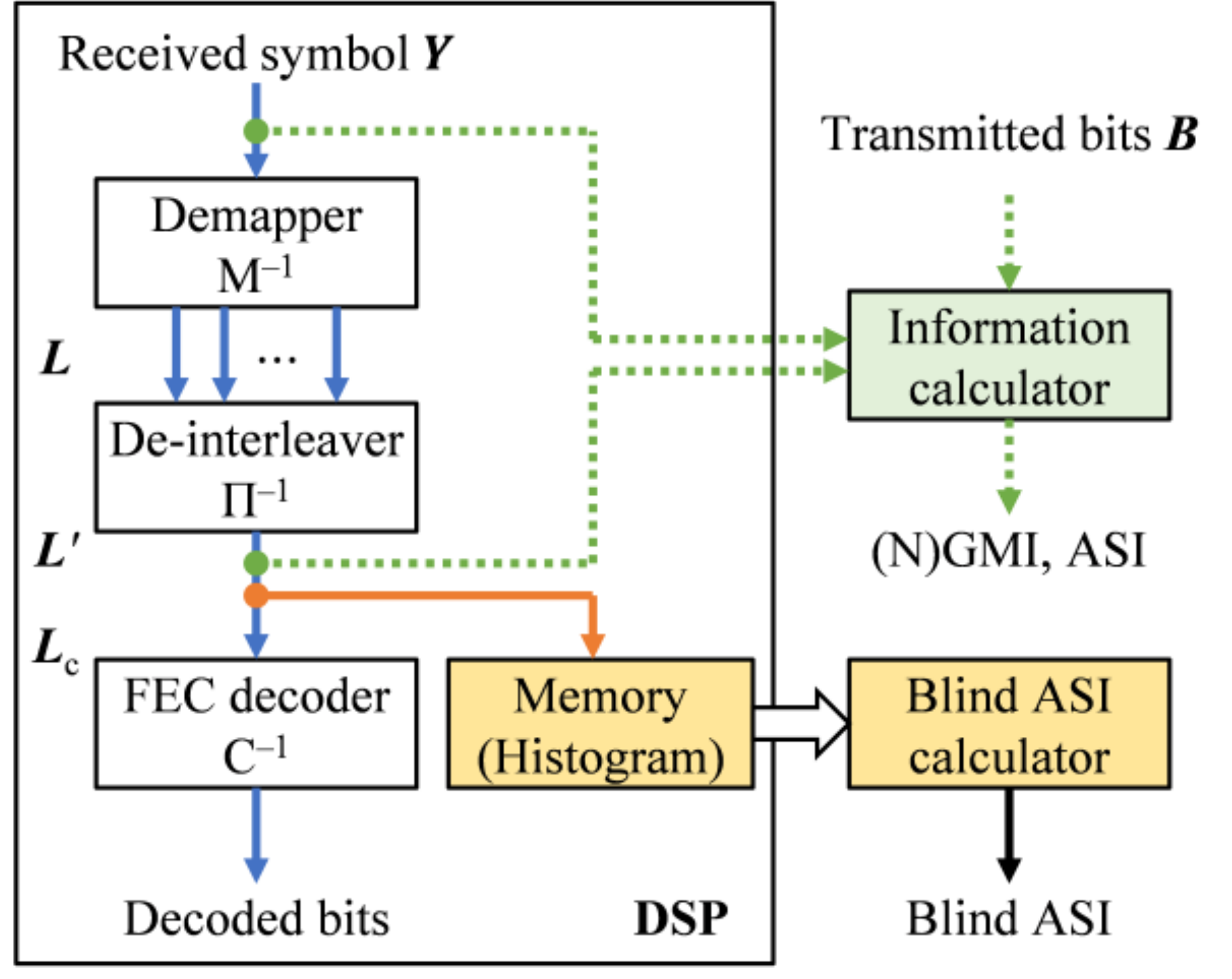} \\
		\vspace{-0.3cm}		
		\caption{Typical computation of information-theoretic performance metrics (green) and functions for performance monitoring of live systems (orange and yellow) around the FEC decoding in DSP.}
		\label{fig:mfnc}
	\end{center}
	\vspace{-0.3cm}
\end{figure}

Fig.~\ref{fig:mfnc} shows the functions around the bitwise FEC decoding in the DSP, which is a part of Fig.~\ref{fig:sys}(c) or \ref{fig:sys}(d). 
This figure explains both the usual derivation of information-theoretic performance metrics and the proposed performance monitoring paths. 
Usually GMI, NGMI, and ASI are computed with the knowledge of the transmitted bits $\boldsymbol{B}$. 
In our proposed technique, we aim to estimate the ASI without this knowledge. We add a memory for storing a histogram of L-values before the FEC decoding, and then extract the histogram to an external function using offline software or a field programmable gate array, and derive a \emph{blind} estimate of the ASI. Although we could have used the received symbol $\boldsymbol{Y}$ instead of $\boldsymbol{L}'$, any approximation or mismatch in the demapping would then not have been accounted for.

\subsection{Principle}
\label{subsec:pri}

\begin{figure}[tb]
	\begin{center}
		\setlength{\unitlength}{.6mm} %
		\scriptsize
		\includegraphics[scale=0.38]{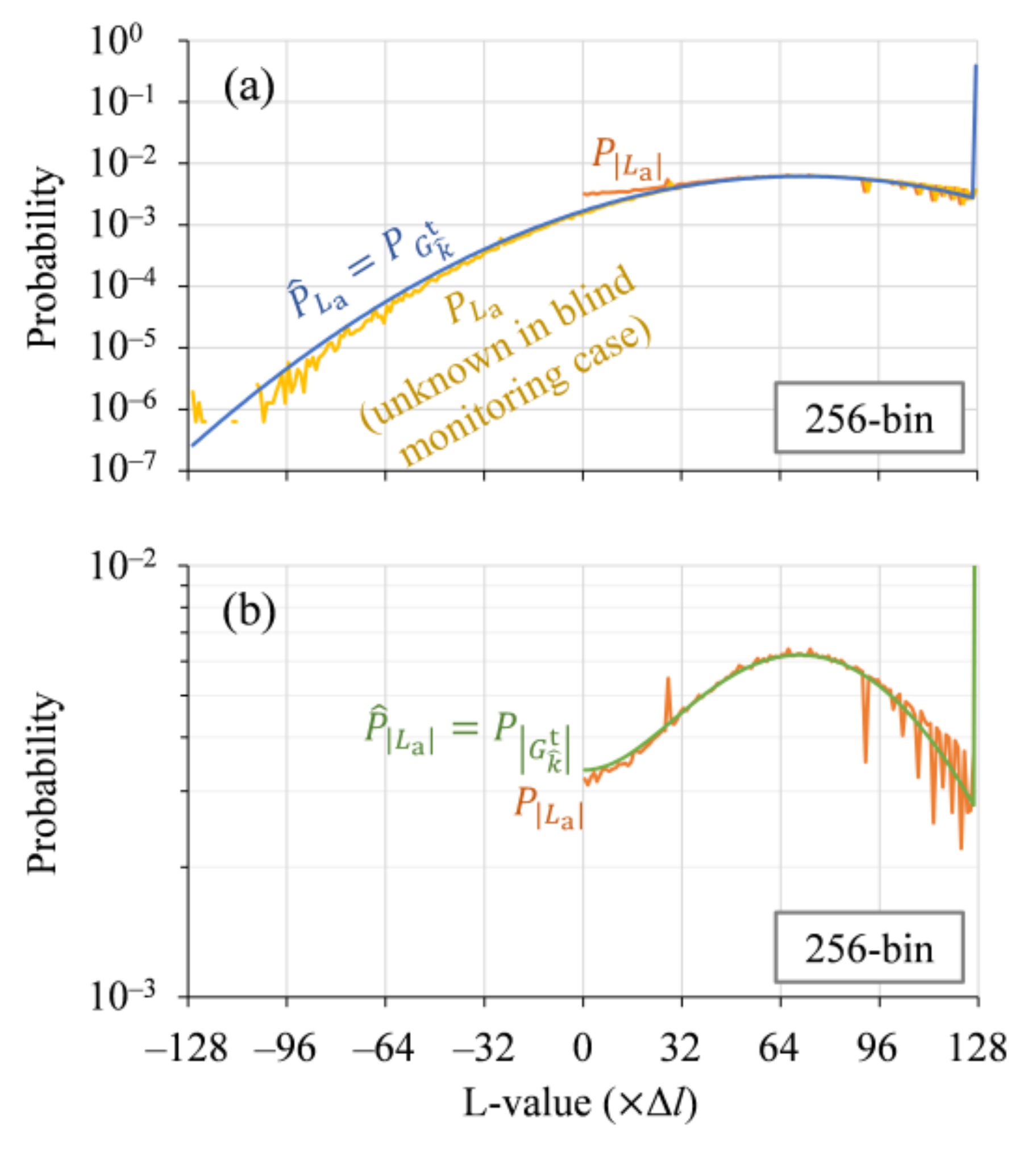} \\
		\vspace{-0.3cm}		
		\caption{Exemplified histogram of L-values for PS-$64$-QAM having $\mathbb{H}(\boldsymbol{B})$ of $4.1\,\mathrm{bpcu}$ and the SNR of $10\,\mathrm{dB}$. The pmfs are as a function of (a) symmetrized \emph{a posteriori} L-value $L_{\mathrm{a}}$ and (b) absolute value $|L_{\mathrm{a}}|$, where $\Delta l =1/13$. \rev{Without the knowledge of the transmitted bits, we know the histogram $P_{|L_{\text{a}}|}(l)$ in (b). In the proposed method, we compare $P_{|L_{\text{a}}|}(l)$ with candidate histograms $P_{|G_k^{\mathrm{t}}|}$ for $k=1,\ldots K$, and choose the best $k$, i.e., $\hat{k}$ in \eqref{eq:full_search}. } }
		\label{fig:Lval}
	\end{center}
	\vspace{-0.3cm}
\end{figure}

Fig.~\ref{fig:Lval} shows an example of an L-value histogram for PS-$64$-QAM at an SNR of $10\,\mathrm{dB}$. Constant-composition distribution matching \cite{schulte_2016} was used to shape the incoming bits into symbols $\boldsymbol{X}$, having a symbol entropy $\mathbb{H}(\boldsymbol{B}) = 4.1\,\text{bits per channel use}\,\mathrm{(bpcu)}$ 
and a PS codeword length $n_\mathrm{ps}=1024$ one-dimensional $8$-ary pulse amplitude modulation symbols. We employed more than $10^6$ samples of the L-values to quantify the performance. 

The ASI can be computed according to \eqref{eq:ASI} if the probability mass function (pmf) $P_{L_{\mathrm{a}}}$ (yellow curve in Fig.~\ref{fig:Lval}(a)) is known. However, under the condition that the transmitted bit $b$ is unknown, the symmetrized L-values $L_{\mathrm{a}}$ are also unknown.
However, the pmf $P_{|L_{\mathrm{a}}|}(l) = P_{|L|}(l)$ (orange curve in Figs.~\ref{fig:Lval}(a) and \ref{fig:Lval}(b)) is known even if $b$ is unknown. Based on real-time measurements of $|L| = |L_{\mathrm{a}}|$, which is quantized to the nearest bin in $\mathbb{L}$ as in \eqref{eq:Lvalue}, we construct a histogram $P_{|L_{\mathrm{a}}|}(l)$, $l \in \mathbb{L}$. 
Due to the finite range of $\mathbb{L}$, there is a peak at $P_{|L_{\mathrm{a}}|}(l_\mathrm{max})$, which represents quantizer overload in \eqref{eq:Lvalue}.

We assume at least the left-side tail of $P_{L_{\mathrm{a}}}$, which mainly determines the ASI, follows a Gaussian distribution.
We therefore precompute a set of discretized Gaussian distributions $P_{G_k^{\mathrm{t}}} (k=1,\ldots ,K)$ and their corresponding symmetrized pmfs $P_{|G_k^{\mathrm{t}}|}$. By comparing $P_{|G_k^{\mathrm{t}}|}$ (green curve in Fig.~\ref{fig:Lval}) with the histogram $P_{|L_{\mathrm{a}}|}$ (orange), the best distribution $P_{G_{\hat{k}}^{\mathrm{t}}}$ is chosen as the estimated pmf $\hat{P}_{L_{\mathrm{a}}}$. The procedure can be summarized as follows:

\begin{enumerate}
\item Select a set of Gaussian distributions\footnote{The constant factor and $P_{G_k}(l_\mathrm{max})$ can be chosen arbitrarily, since these will not be used in \eqref{eq:pgkt}--\eqref{eq:hatQstLval}.} $f_k(l) \triangleq \exp[-(l-\mu[k])^2/(2\sigma^2[k])]$ for $k=1, \ldots , K$ and discretize them into
\begin{align}
	P_{G_k}(l) &\triangleq
	\begin{cases}
		\int_{l-\Delta l/2}^{l+\Delta l/2} f_k(l) \mathrm{d}l, & l \in \mathbb{L}\setminus \{-l_\mathrm{max},l_\mathrm{max}\}, \\
		\int_{-\infty}^{-l_{\mathrm{max}}+\Delta l/2} f_k(l) \mathrm{d}l, & l = -l_{\mathrm{max}}.
	\end{cases}
\end{align}
\item Set $\rho \triangleq P_{|L_{\mathrm{a}}|}(l_{\mathrm{max}})$.
\item For $k=1,2,\ldots,K$, let 
\begin{align}
	\label{eq:pgkt}
	P_{G_k^{\mathrm{t}}}(l) &\triangleq
	\begin{cases}
		\frac{1-\rho}{\sum_{l' \in \mathbb{L}\setminus \{l_\mathrm{max}\}} P_{G_k}(l')} P_{G_k}(l), & l \in \mathbb{L}\setminus \{l_\mathrm{max}\}, \\
		\rho, & l = l_{\mathrm{max}},
	\end{cases}
\end{align}
where the normalization serves to satisfy $\sum_{l\in\mathbb{L}}P_{G_k^{\mathrm{t}}}(l)=1$, and calculate $P_{|G_k^{\mathrm{t}}|}(l) = P_{G_k^{\mathrm{t}}}(-l) +P_{G_k^{\mathrm{t}}}(l)$ for $l \in \mathbb{L}^+ = \{l \in \mathbb{L}: l>0\}$.
\item Find by full search
\begin{IEEEeqnarray}{rCL}
	\label{eq:full_search}
	\hat{k} \triangleq \argmin_k \sum_{l\in \mathbb{L}^+} \left( P_{|L_{\mathrm{a}}|}(l) - P_{|G_k^{\mathrm{t}}|}(l) \right) ^2 .
\end{IEEEeqnarray}
Other optimization methods, e.g., gradient descent over the distribution parameters, can be employed instead.
\item Assume that the parameters $\mu[\hat{k}]$ and $\sigma[\hat{k}]$ estimated for $|L| = |L_{\mathrm{a}}|$ also are valid for $L_{\mathrm{a}}$; i.e., set 
\begin{IEEEeqnarray}{rCL}
	\hat{P}_{L_{\mathrm{a}}}(l) \triangleq P_{G_{\hat{k}}^{\mathrm{t}}}(l),\quad l \in \mathbb{L}.
\end{IEEEeqnarray}
\item Calculate an estimate of the ASI, called \emph{blind} ASI and denoted \rev{$\hat{I}_{\mathrm{a}}$} in the remains of this paper, by \eqref{eq:ASI}, using $\hat{P}_{L_{\mathrm{a}}}$ 
to approximate $P_{L_{\mathrm{a}}}$.
\item Optionally, an estimate of the Q-factor
\begin{IEEEeqnarray}{rCL}
	\label{eq:hatQstLval}
	\rev{\hat{Q}_{\mathrm{st,L}}} & \rev{\, = \,} & \rev{ \mathrm{erfc}^{-1} \left(2 ( 1 - \rho ) \widehat{BER}_{\mathrm{Gauss}} \right) }
\end{IEEEeqnarray}
can be computed as a by-product\rev{, where,}
\begin{IEEEeqnarray}{rCL}
	\label{eq:hatQstLval}
	\rev{\widehat{BER}_{\mathrm{Gauss}}} & \rev{\, = \,} & \rev{ \frac{1}{2} \mathrm{erfc} \left( \frac{\mu[\hat{k}]}{\sqrt{2} \sigma[\hat{k}]} \right) .}
\end{IEEEeqnarray}
\end{enumerate}
In \eqref{eq:hatQstLval}, the factor $(1 - \rho)$ removes the influence of the peak due to quantization overload in the histogram of L-values, which we did not consider in \cite{yoshida_2019_ecoc}.

\subsection{Initial simulation}
\label{subsec:sim}

\begin{figure}[tb]
	\begin{center}
		\setlength{\unitlength}{.6mm} %
		\scriptsize
		\includegraphics[scale=0.43]{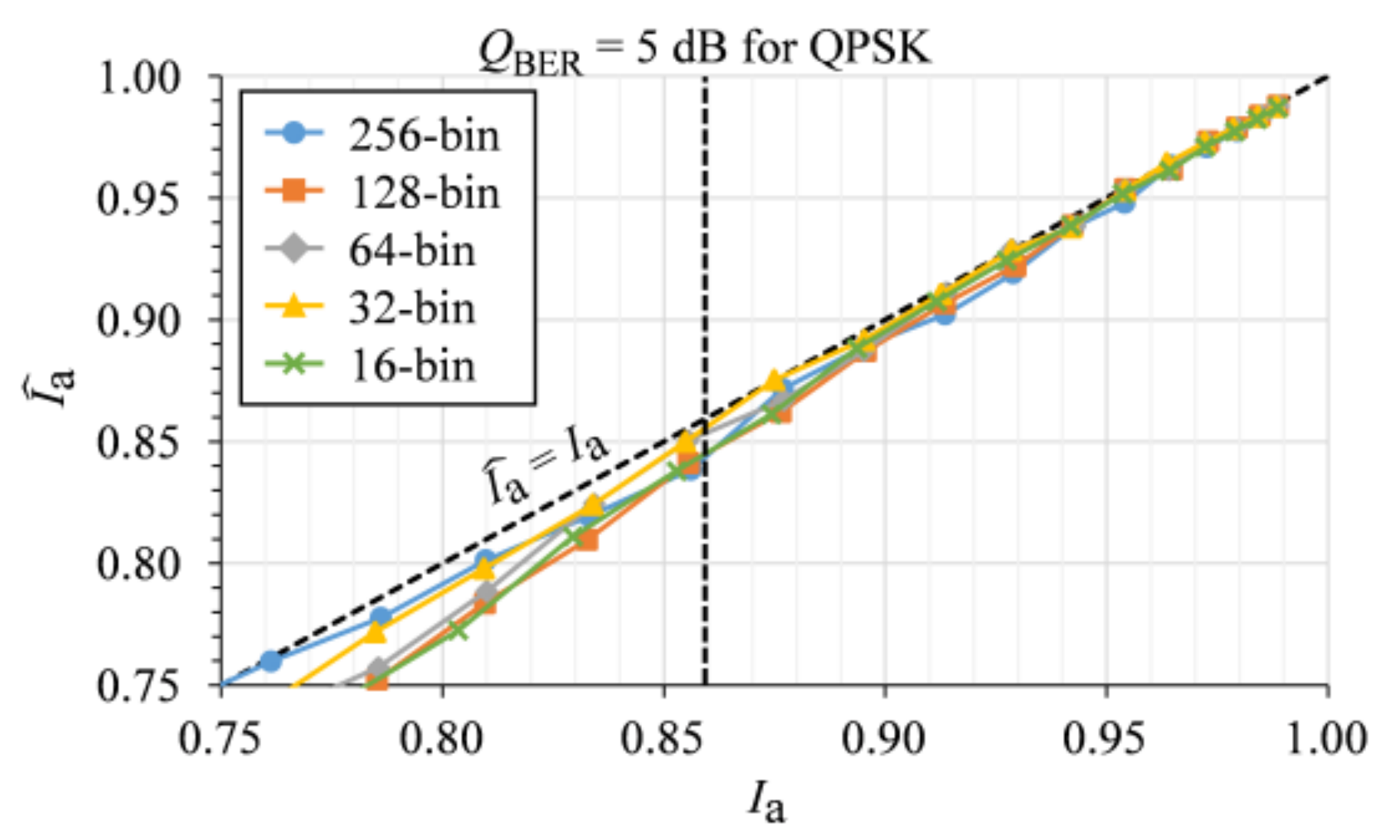} \\
		\vspace{-0.3cm}		
		\caption{Performance of ASI estimation for different $n_\mathrm{bin}$ values for PS-$64$-QAM having a symbol entropy $\mathbb{H}(\boldsymbol{B})$ of $4.1\,\mathrm{bpcu}$ over the Gaussian channel.}
		\label{fig:sim}
	\end{center}
	\vspace{-0.3cm}
\end{figure}

In Fig.~\ref{fig:Lval}, \rev{the} ASI from the sampled $P_{|L_{\mathrm{a}}|}$ using knowledge of the transmitted bits, \rev{$I_{\mathrm{a}}$}, is $0.876$ and that from the proposed estimation method, \rev{$\hat{I}_{\mathrm{a}}$}, is $0.870$, so the absolute estimation error is $0.006$.
Fig.~\ref{fig:sim} shows the absolute estimation errors of the ASI by the proposed method with different resolutions of the L-value histogram from $n_\mathrm{bin} =16$ to $256\,\mathrm{bins}$ ($4$ to $8\,\mathrm{bits}$). \rev{The resolution is the same as for the L-values fed into the soft FEC decoder.}
The \rev{$I_{\mathrm{a}}$} was changed along the horizontal axis by sweeping the SNR from $5\,\mathrm{dB}$ to $15\,\mathrm{dB}$. 
The SNR of the auxiliary channel was set to a\rev{n} FEC threshold condition of $\rev{I_{\mathrm{a}}}=0.86$\cite{sugihara_2013}, which corresponds to a $Q_{\mathrm{BER}}$ of $5\,\mathrm{dB}$ for QPSK over the Gaussian channel. 
\rev{Note that there are random fluctuations between simulation instances, and that a coarse quantization with a small number of bins degrades the signal quality of \rev{$I_{\mathrm{a}}$}. With at least $32\,\mathrm{bins}$, the reduction is negligible. 
In Fig.~\ref{fig:Lval}, in the regime where \rev{$I_{\mathrm{a}}$} is more than the FEC threshold (here 0.86) is much more important than the others for the system margin estimation. 
In this regime, the estimation errors for the examined bin sizes (with fluctuations between simulation instances) are not significantly different. }
The absolute \rev{estimation} error\rev{s} \sout{in the estimation is} \rev{are} \sout{then} within $0.015$ \sout{for ASI values above the FEC threshold of $0.86$}, giving a relative error of $1.7\%$. 

When computing $\mathbb{H}( L_{\mathrm{a}} \mid |L_{\mathrm{a}}| )$ in \eqref{eq:ASI} in step 6, $P_{|L_{\mathrm{a}}|}$ can be used instead of $\hat{P}_{|L_{\mathrm{a}}|}$ to compute \rev{$\hat{I}_{\mathrm{a}}$}. In our simulations, we did not see any significant difference between these two cases.

\section{Experiments}
\label{sec:exp}

We examined the estimation accuracy of the proposed blind performance monitoring method over fiber-optic channels by recirculating loop transmission and back-to-back noise loading experiments. 

\subsection{Transmission experiments}
\begin{figure}[tb]
	\begin{center}
		\setlength{\unitlength}{.6mm} %
		\scriptsize
		\includegraphics[scale=0.8]{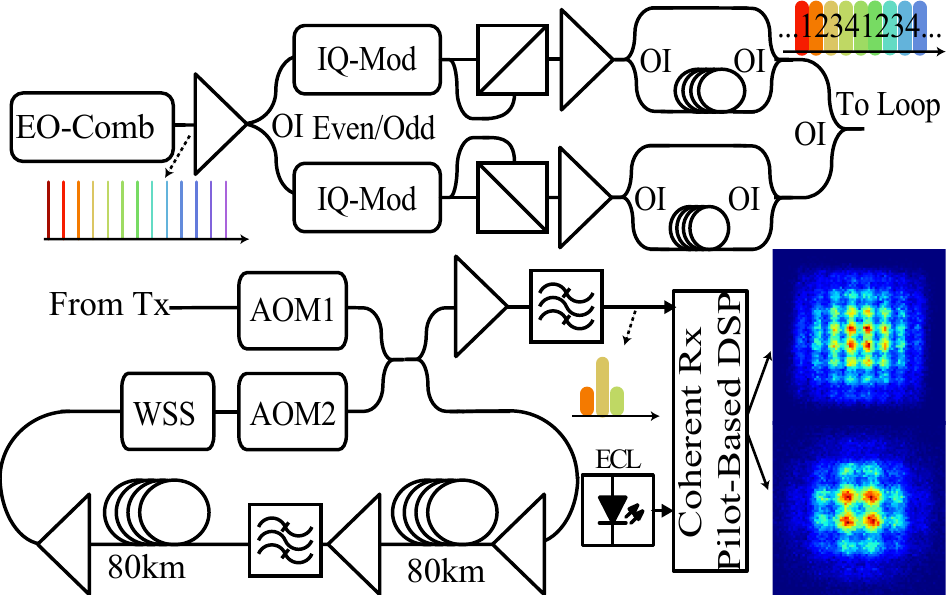} \\
		\vspace{-0.3cm}		
		\caption{Experimental setup. Insets are recovered constellations for
      PS-64-QAM in the case of $\mathbb{H}(\boldsymbol{B})=5.7\,\mathrm{bpcu}$
      after $5$ roundtrips (top) and
      $\mathbb{H}(\boldsymbol{B})=4.6\,\mathrm{bpcu}$ after $13$ roundtrips (bottom). }
		\label{fig:exp_setup}
	\end{center}
	\vspace{-0.3cm}
\end{figure}

The experimental setup is shown in Fig.~\ref{fig:exp_setup}. At the transmitter,
we used an electro-optic frequency comb with $51$ tones spaced at $25\,\mathrm{GHz}$ seeded by
an external cavity laser (ECL).
The lines were separated by an optical interleaver (OI) into even and odd
channels which were each modulated by an in-phase and quadrature modulator
(IQ-Mod) driven by a $60\,\mathrm{GS/s}$ arbitrary waveform
generator. The symbol rate of the channels was $24\,\mathrm{Gbaud}$. We emulated polarization division
multiplexing by a split-delay-combine technique with about $250$ symbols delay.
Every second of the dual-polarization odd and even channels was further
decorrelated
with anther set of OIs and
decorrelation fibers corresponding to about $750$ symbols delay to generate a
$1-2-3-4-1-2\dots$ decorrelation pattern. All
channels were then multiplexed and coupled into a recirculating loop using
two acusto-optic modulators (AOM). The loop contained two spans of $80\,\mathrm{km}$ standard single-mode fiber (SSMF) and
the \rev{default} total input power was $13\,\mathrm{dBm}$, corresponding to a per-channel launch power of
$-4\,\mathrm{dBm}$.
\rev{We examined the launch power dependence under limited conditions before sweeping the number of roundtrips, and we observed that $-4\,\mathrm{dBm}$ is optimal for $5$--$20$ roundtrips ($800$--$3200\,\mathrm{km}$). With the dispersion-uncompensated link in this experiment, the optimal launch power is almost the same even if the number of spans differs. With this default launch power, a moderate amount of fiber nonlinearity influenced the signal quality.}
Three in-loop erbium-doped fiber amplifiers (EDFAs)
with a noise figure of approximately $5\,\mathrm{dB}$ compensated for fibre losses and an optical bandpass filter and a
wavelength selective switch were employed for rejection of
out-of-band ASE noise and compensation of gain tilt.
At the receiver the signal was coherently detected by mixing with a local oscillator from
another ECL. Receiver side DSP was performed by offline processing; after
static chromatic dispersion compensation, we performed fully pilot-aided signal recovery
\cite{mazur_2019}.\footnote{\rev{Alternatively, one can estimate the signal quality based on pilot symbols. 
However, such estimates will not be representative for the whole signal, since pilot symbols usually have a low-order modulation format (in our case, QPSK) without FEC coding, whereas higher-order modulation and FEC are applied to data symbols. Furthermore, the accuracy of pilot-based signal quality estimates is relatively low, since the pilot insertion ratio is usually limited to a few \% in deployable systems not to reduce the throughput of the main signal. 
}}
The adaptive equalization consisted of butterfly-structured finite-impulse-response filters with $45$ half-symbol taps based on
the constant modulus algorithm of the pilot sequence with a length of $2048$
symbols.
The carrier phase was estimated by linear interpolation of the phases of
interleaved pilot symbols (one per 64 symbols).
Examples of recovered constellations are shown in the insets of Fig.~\ref{fig:exp_setup}.

We investigated $16$-QAM, $64$-QAM, PS-$16$-QAM, PS-$64$-QAM, and PS-$256$-QAM. PS-$64$-QAM
signals with
$\mathbb{H}(\boldsymbol{B})=4.1,\,4.6,\,5.2,$ and $5.7\,\mathrm{bpcu}$
were generated by CCDM \cite{schulte_2016}. PS-$16$-QAM and PS-$256$-QAM signals
were generated with the PS coding method from \cite{yoshida_2018_ofc_spg} for $\mathbb{H}(\boldsymbol{B})= 3.4$\rev{, $5.0$,} and $6.3\,\mathrm{bpcu}$, resp\rev{ectively}. 
The SNR of the auxiliary channel was set to a pre-set value per
modulation/shaping format at a\rev{n} \sout{typical} FEC threshold ASI of \rev{FT\#1: $0.93$, FT\#2: $0.86$, or FT\#3: $0.78$} regardless of the number of roundtrips. The number of
L-value bins $n_{\mathrm{bin}}$ was set to $32$ as a reasonable value for practical DSP.
\rev{To observe one histogram of L-values, we stored $10^5$ polarization-multiplexed QAM symbols.}
The number of examined Gaussian distributions $K$ in the proposed scheme was $8192$ (see Sec.~\ref{subsec:pri}, step 1).

\begin{figure}[t]
	\begin{center}
		\setlength{\unitlength}{.6mm} %
		\scriptsize
		\includegraphics[scale=0.43]{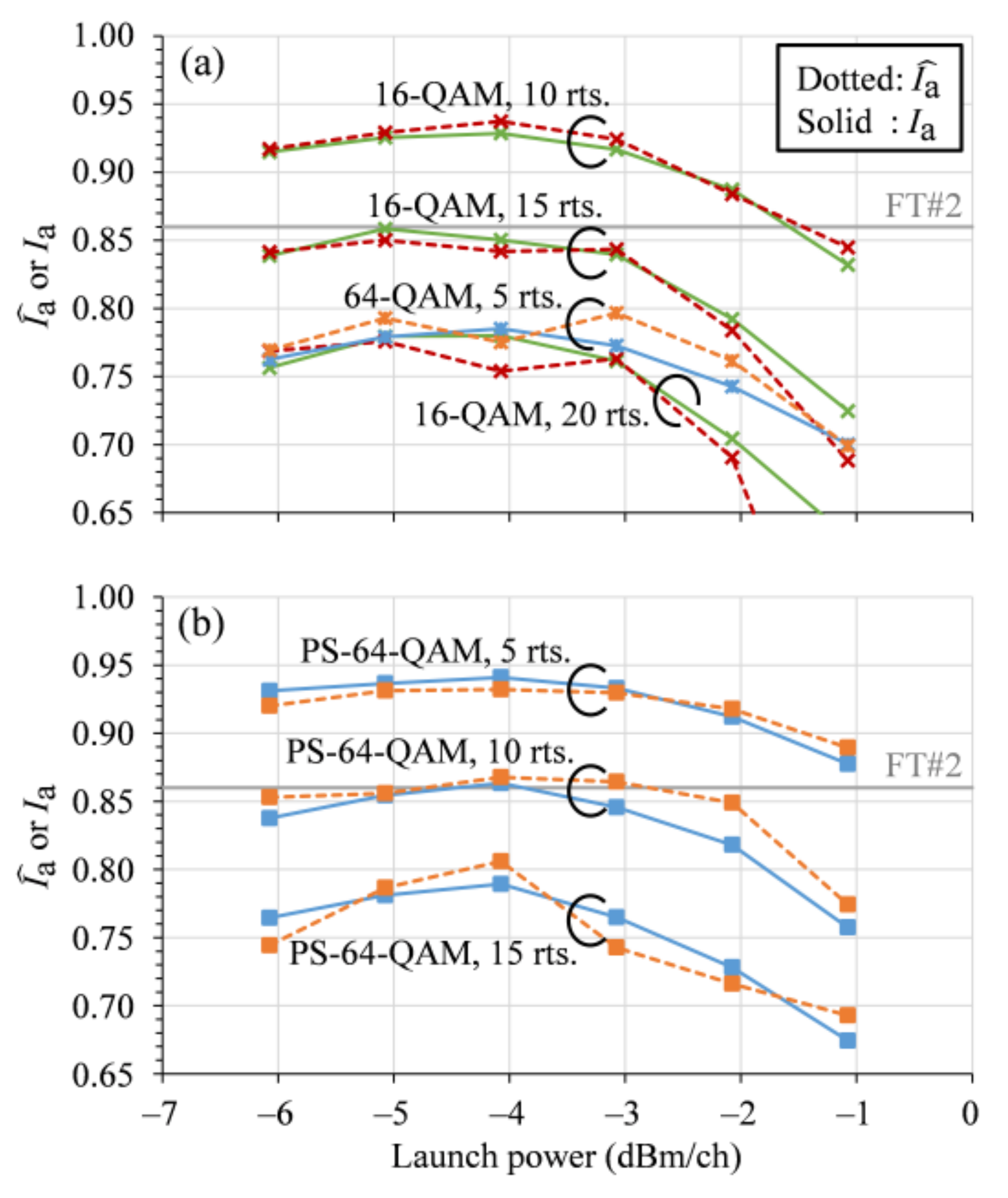} \\
		\vspace{-0.3cm}
		\caption{\rev{True and estimated ASI in in transmission experiments with varying launch power for (a) uniform $16$-QAM ($10$, $15$, and $20$ roundtrips) and $64$-QAM ($5$ roundtrips), and (b) PS-$64$-QAM with $\mathbb{H}(\boldsymbol{B})$ of $5.0\,\mathrm{bpcu}$ ($5$, $10$, and $15$ roundtrips).}}
		\label{fig:exp_lp}
	\end{center}
	\vspace{-0.3cm}
\end{figure}

\rev{Fig.~\ref{fig:exp_lp} shows the ASI as a function of the launch power based on knowledge of the transmitted bits (solid lines, \rev{$I_{\mathrm{a}}$}) and with the proposed
method without such knowledge (dotted lines, \rev{$\hat{I}_{\mathrm{a}}$}) in selected cases. The modulation/shaping parameters were $16$-QAM, PS-$64$-QAM with $\mathbb{H}(\boldsymbol{B})=5.0\,\mathrm{bpcu}$, and $64$-QAM. The assumed FEC threshold was FT\#2.
In the examined cases, around $-4\,\mathrm{dBm/ch}$ is optimal. Except for the case of very low performance of $16$-QAM with $20$ roundtrips and $-1\,\mathrm{dBm/ch}$, the maximum absolute estimation error is $0.036$ for $16$-QAM with $15$ roundtrips and $-1\,\mathrm{dBm/ch}$. Above the assumed FEC threshold FT\#2, the estimation errors are much smaller.}

\begin{figure}[t]
	\begin{center}
		\setlength{\unitlength}{.6mm} %
		\scriptsize
		\includegraphics[scale=0.43]{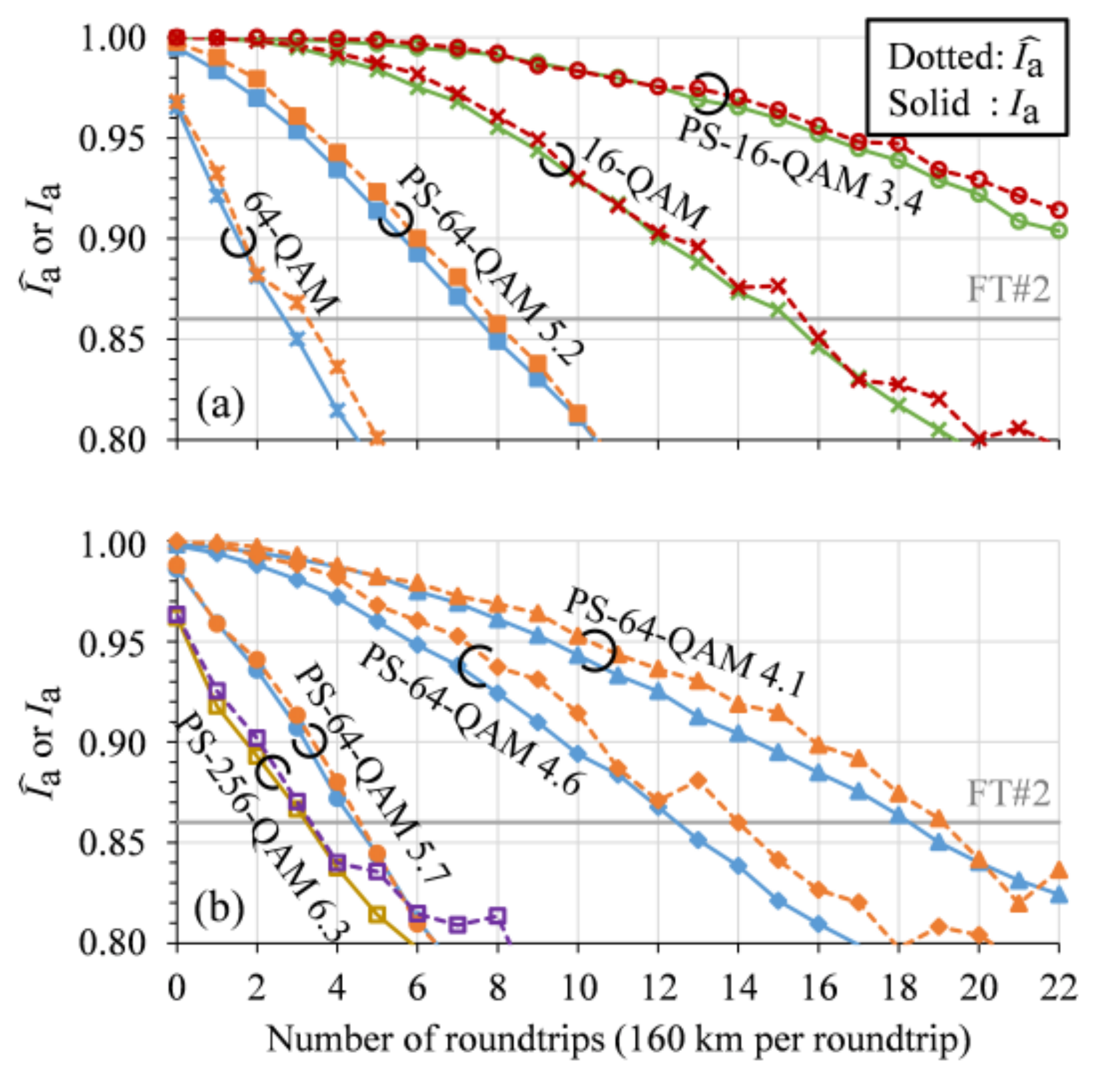} \\
		\vspace{-0.3cm}
		\caption{True and estimated ASI in transmission experiments \rev{at a launch power of $-4\,\mathrm{dBm/ch}$ for (a) PS-$16$-QAM ($\mathbb{H}(\boldsymbol{B})=3.4\,\mathrm{bpcu}$), $16$-QAM, PS-$64$-QAM ($\mathbb{H}(\boldsymbol{B})=5.2\,\mathrm{bpcu}$), and $64$-QAM, and (b) PS-$64$-QAMs ($\mathbb{H}(\boldsymbol{B})=4.1$, $4.6$, and $5.7\,\mathrm{bpcu}$), and PS-$256$-QAM ($\mathbb{H}(\boldsymbol{B})=6.3\,\mathrm{bpcu}$).}}
		\label{fig:exp_asi}
	\end{center}
	\vspace{-0.3cm}
\end{figure}

Fig.~\ref{fig:exp_asi} shows the ASI as a function of roundtrips
\sout{based on knowledge of the transmitted bits (solid lines) and with the proposed
method without such knowledge (dotted lines)} \rev{for eight modulation/shaping parameters under FT\#2 and a launch power of $-4\,\mathrm{dBm/ch}$}.
\sout{The markers for base constellations $16$, $64$, and $256$-QAMs are colored
green/red, blue/orange, and yellow/purple, resp\rev{ectively}.}
We can see that the \rev{$I_{\mathrm{a}}$} and \rev{$\hat{I}_{\mathrm{a}}$} are well correlated, although \rev{$\hat{I}_{\mathrm{a}}$} tends to be \sout{slightly} larger than the true values \rev{in the regime shown in Fig.~\ref{fig:exp_asi}}.

\begin{figure}[tb]
	\begin{center}
		\setlength{\unitlength}{.6mm} %
		\scriptsize
		\includegraphics[scale=0.43]{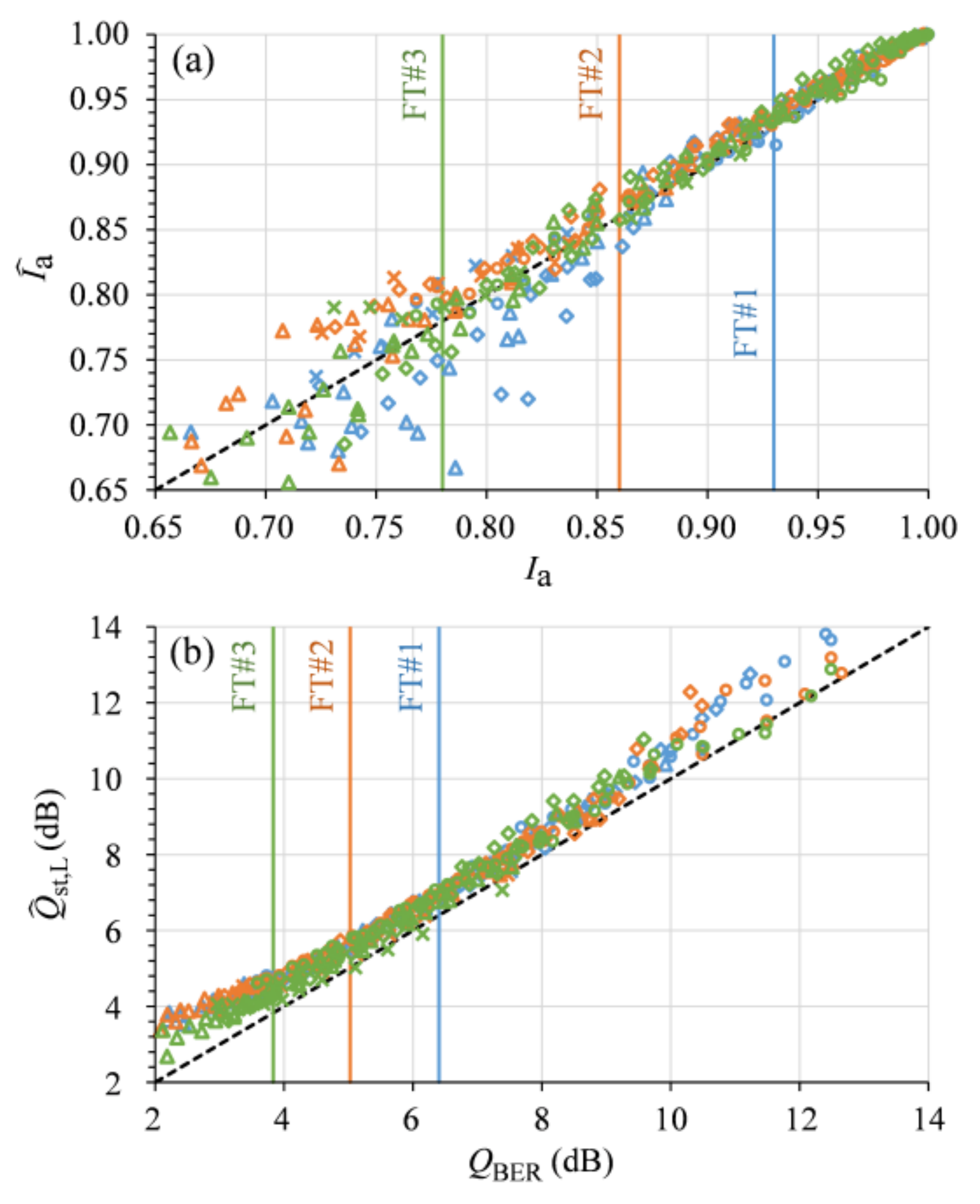} \\
		\vspace{-0.3cm}		
		\caption{Estimation errors \rev{with the proposed method in} (a) \rev{soft (}ASI\rev{)} and (b) \rev{hard (}Q-factor\rev{) performance monitoring} in transmission experiments. \rev{Assumed FEC threshold is FT\#1: $0.93$ (blue), FT\#2: $0.86$ (orange), or FT\#3: $0.78$ (green). The corresponding Q-factors with binary modulation are
		$6.4\,\mathrm{dB}$, $5.0\,\mathrm{dB}$, or $3.8\,\mathrm{dB}$, respectively. Circles, diamonds, triangles, or crosses depict the cases of $16$-QAM and PS-$16$-QAM, PS-$64$-QAM with $\mathbb{H}(\boldsymbol{B}) \le 5.0\,\mathrm{bpcu}$, PS-$64$-QAM with $\mathbb{H}(\boldsymbol{B}) > 5.0\,\mathrm{bpcu}$ and 64-QAM, or PS-$256$-QAM, respectively. The dotted line indicates where the estimated value equals to the true value.}}
		\label{fig:exp_err_trans}
	\end{center}
	\vspace{-0.3cm}
\end{figure}

Fig.~\ref{fig:exp_err_trans} shows the \sout{absolute estimation errors} \rev{relationship between the true and  estimated values of} \sout{in} (a) ASI \sout{and} \rev{or} (b) Q-factor \rev{under the assumed FEC threshold ASIs.} \sout{versus true ASI and Q-factor resp\rev{ectively}.} 
\rev{Note that the Q-factor is computed by \eqref{eq:hatQstLval}, which is a byproduct of the proposed ASI estimation method.}
In Fig.~\ref{fig:exp_err_trans}(a), 
\rev{
the estimation in the regime where $I_{\mathrm{a}}$ is larger than or equal to the assumed FEC threshold, the estimation errors are small.
The maximum absolute (and relative) estimation errors are $0.016$ ($1.7\%$), $0.021$ ($2.3\%$), and $0.028$ ($3.6\%$) for FT\#1, FT\#2, and FT\#3, respectively.
}
\sout{for $\mathit{ASI}>0.8$, the absolute error is
less than $0.03$ and the relative error is less than $3.5\%$.
Above the assumed FEC threshold of $\mathit{ASI}=0.86$, the errors are further
reduced to less than $0.022$ and less than $2.4\%$ for absolute and relative error
resp\rev{ectively}, which}
\rev{These estimation errors} correspond\sout{s} to \sout{around} \rev{within} $0.5\,\mathrm{dB}$ SNR difference for binary
modulation over the Gaussian channel.
A reason for the positively biased errors \rev{above each FEC threshold} in experiments could be the larger
gap between the L-value and assumed Gaussian distributions.
The L-value distribution for the fiber-optic channel could be different from
that of the Gaussian channel due to imperfections of the optical transceiver and
the DSP.
In the DSP, normalization and quantization around soft demapping will influence the blind estimation results.
We observe that the estimation error becomes larger for smaller \rev{$I_{\mathrm{a}}$},
especially below the assumed FEC limit.
We believe this is due to the fact that the true SNR is significantly smaller
than the SNR of the auxiliary channel.
For example, the SNR of the auxiliary channel was $3\,\mathrm{dB}$ larger than the
approximated true SNR in the case of PS-256-QAM after $8$ roundtrips.
\rev{In such cases, the mismatched L-values are significantly larger than the true L-values, which would influence the estimation error.}

We have also tried to estimate the ASI per bit tributary separately and averaged them to obtain a bit-averaged ASI, however this did not improve the estimation accuracy.

Fig.~\ref{fig:exp_err_trans}(b) depicts the \rev{by-product} Q-factor estimation error\rev{.} \sout{$\Delta_Q\,\mathrm{(dB)} =
\hat{Q}_{\mathrm{st}}\,\mathrm{(dB)} - Q_{\mathrm{BER}}\,\mathrm{(dB)}$ where the
estimated Q-factor $\hat{Q}_{\mathrm{st}}$ is derived from \eqref{eq:hatQstLval}. } 
Similar to the ASI estimate we see a positive bias of the estimated Q-factor
$\hat{Q}_{\mathrm{st\rev{,L}}}$.
The peak-to-peak \sout{$\Delta_Q$} \rev{estimation error in the Q-factor} \sout{variation} \rev{is within $1.1\,\mathrm{dB}$\rev{, $0.8\,\mathrm{dB}$, or $1.5\,\mathrm{dB}$ at a} $Q_{\mathrm{BER}}$} \sout{of $5$--$9\,\mathrm{dB}$} \rev{larger than each assumed FEC threshold and lower than $9\,\mathrm{dB}$ (i.e., $3.8$--$9.0\,\mathrm{dB}$ for FT\#1, $5.0$--$9.0\,\mathrm{dB}$ for FT\#2, or $6.4$--$9.0\,\mathrm{dB}$ for FT\#3, respectively).} \sout{except for a larger extrapolated gap at $Q_{\mathrm{BER}}=5\,\mathrm{dB}$ for
PS-16-QAM ($\mathbb{H}(\boldsymbol{B})=3.4\,\mathrm{bpcu}$).}
By introducing the correction term $(1-\rho)$ in \eqref{eq:hatQstLval}, the
peak-to-peak \sout{$\Delta_Q$} \rev{estimation error in the Q-factor} \sout{variation} is almost halved. 
We also observe that \sout{$\Delta_Q$} $\hat{Q}_{\mathrm{st,L}} - Q_{\mathrm{BER}}$ depends on the symbol entropy $\mathbb{H}(\boldsymbol{B})$, i.e., a \sout{small} \rev{larger} entropy signal \rev{such as PS-$256$-QAM (crosses in the figure)} tends to have \sout{larger} \rev{smaller} \sout{$\Delta_{Q}$} \rev{$\hat{Q}_{\mathrm{st,L}} - Q_{\mathrm{BER}}$}. \sout{PS-$16$-QAM ($\mathbb{H}(\boldsymbol{B})=3.4\,\mathrm{bpcu}$) exhibits the largest and PS-256-QAM ($\mathbb{H}(\boldsymbol{B})=6.3\,\mathrm{bpcu}$) the smallest $\Delta_Q$ for $Q_{\mathrm{BER}}$ just above the assumed FEC threshold.} This dependence was not found in the ASI estimation.

\subsection{Back-to-back experiments}
\begin{figure}[tb]
	\begin{center}
		\setlength{\unitlength}{.6mm} %
		\scriptsize
		\includegraphics[scale=0.43]{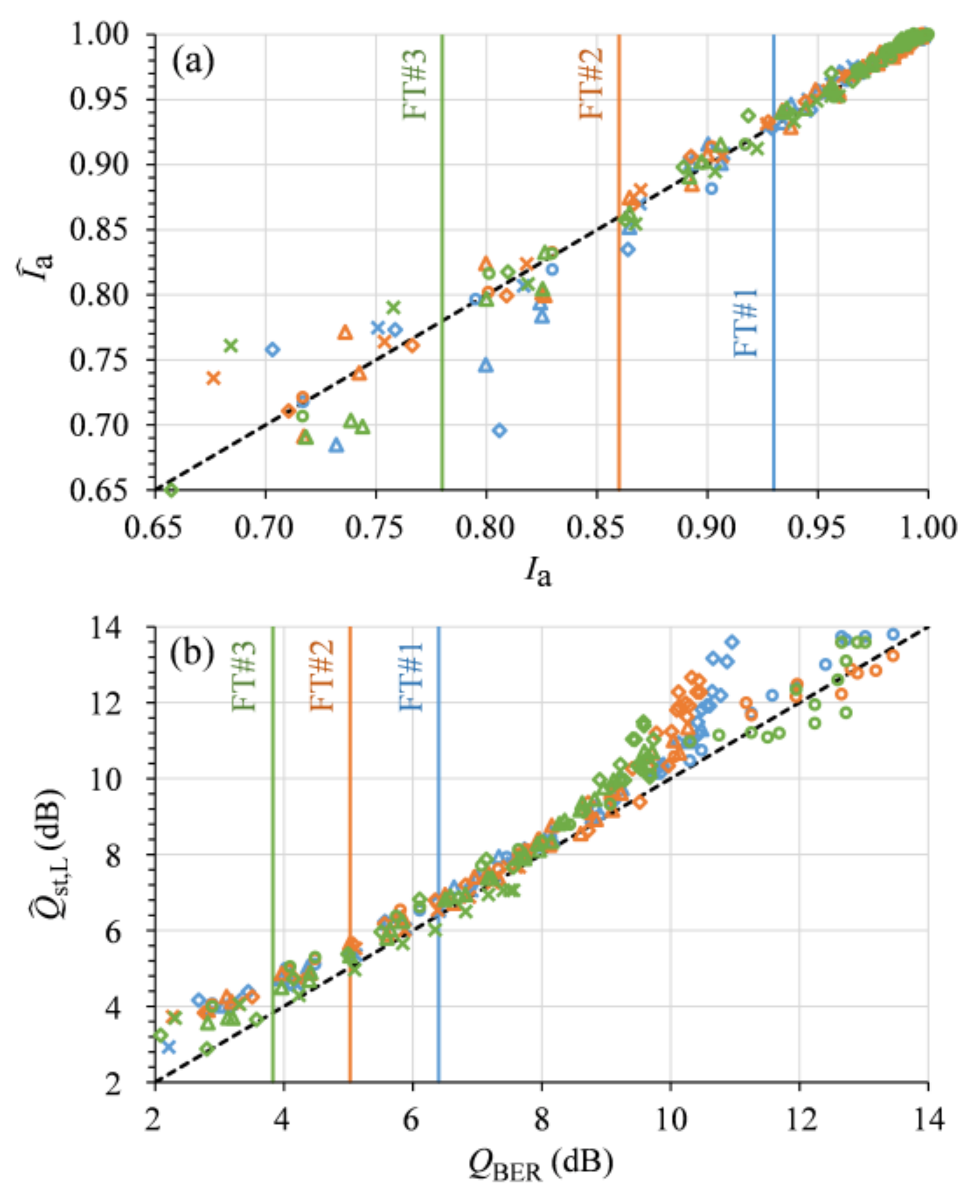} \\
		\vspace{-0.3cm}		
		\caption{Estimation errors \rev{with the proposed method} in (a) \rev{soft (}ASI\rev{)} and (b) \rev{hard (}Q-factor\rev{) performance monitoring} in B2B noise loading experiments. \rev{Legends and colors are the same as in Fig.~\ref{fig:exp_err_trans}.}}
		\label{fig:exp_err_btb}
	\end{center}
	\vspace{-0.3cm}
\end{figure}

To confirm if the trends observed in the transmission experiments are specific
to transmission, we captured and processed data in a back-to-back, noise loading
configuration. We swept the OSNR from $10$ to $36\,\mathrm{dB}$ at $0.1\,\mathrm{nm}$ noise bandwidth.
Fig.~\ref{fig:exp_err_btb} show the estimation errors of (a) ASI and (b)
\rev{by-product} Q-factor.
In Fig.~\ref{fig:exp_err_btb}(a), we see \rev{no} positive bias of \rev{$\hat{I}_{\mathrm{a}}$}
over \rev{$I_{\mathrm{a}}$}, while the absolute estimation error is comparable with the
transmission experiments.
In Fig.~\ref{fig:exp_err_btb}(b), the back-to-back estimated Q-factor exhibits a
similar behaviour as in the transmission experiments, i.e., positively bias
compared to the true $Q_{\mathrm{BER}}$. The estimation errors are within $1\,\mathrm{dB}$ at
$Q_{\mathrm{BER}}$ \sout{$= 5$--} \rev{of each FEC threshold to} $9\,\mathrm{dB}$,
which is almost the same as in the transmission experiments.
For pure noise-limited situations there would be no specific difference in
L-value histograms between the transmission and noise loading cases. However, in
the transmission case fiber nonlinearities could 
cause distribution differences which influences the ASI. 

When the ASI is close to 1, the ASI is a highly nonlinear function of SNR (although $\log
(-\log (1-\rev{I_{\mathrm{a}}}))$ versus SNR is somewhat more linear
\cite[Fig.~3]{yoshida_2019_arxiv}). This causes the ASI estimation error
to approach zero around $\rev{I_{\mathrm{a}}}=1$. It can therefore be said that ASI or
other information theoretric metrics are not suitable to quantify the signal
performance in this high performnace regime, e.g. at SNRs corresponding to
$Q_{\mathrm{BER}}=10\,\mathrm{dB}$. On the other hand, the Q-factor is well suited for this
regime, even though the estimation error using our proposed method becomes larger.

\section{Conclusions}
\label{sec:cnc}
We proposed a blind performance monitoring method for systems with soft binary
FEC and multilevel modulation. The proposed method estimates the ASI based on
a histogram of the \emph{a posteriori} L-values without any knowledge of the
transmitted bits.
We compare our monitored L-values to a set of Gaussian distribution candidates to approximate them by a Gaussian distribution.

\sout{Eight} \rev{Nine} different modulation/shaping formats \rev{and three assumed FEC thresholds} were examined by middle- to long-haul
transmission experiments\rev{.} \sout{and} \rev{W}e observed positively biased absolute errors,
with a maximum estimation error below \rev{$0.016$ ($1.7\%$), $0.021$ ($2.3\%$), or $0.028$ ($3.6\%$)} for ASIs larger than the assumed FEC
threshold of $\rev{I_{\mathrm{a}}} = \rev{0.78}$\rev{, }$0.86$\rev{, or $0.93$, respectively}.
This estimation error corresponds to \rev{within} $0.5\,\mathrm{dB}$ SNR difference for binary modulation in the Gaussian channel in the regime of $\rev{I_{\mathrm{a}}}=0.\rev{78}$--$0.9\rev{3}$. 
As a by\rev{-}product of the proposed method, we estimated a hard decision
performance metric of $Q_{\mathrm{BER}}$. The estimation errors were also
positively biased with a maximum absolute error less than $\rev{0.8}\,\mathrm{dB}$. 
Finally we investigated the estimation errors with back-to-back noise loading
experiments to remove the effects from transmission.
Only the ASI exhibits a difference between the two back-to-back and transmission experiments,
i.e., the estimated ASI was not positively biased in back-to-back noise loading
experiments. Although the mechanism of the estimation bias is not fully
clarified, it is likely that the nonlinear fiber transmission changes the
L-value distribution away from a Gaussian distribution, inducing an ASI
estimation bias.
Our proposed method is well suited for performance monitoring in systems
carrying live traffic, because only minimal additional functionality is required
in the hardware that
stores histograms of the L-values before FEC decoding.

\section*{Acknowledgement}
We thank Assoc. Prof. Koji Igarashi of Osaka University for fruitful discussions.

\balance


\begin{thebibliography}{10}
	\providecommand{\url}[1]{#1}
	\csname url@samestyle\endcsname
	\providecommand{\newblock}{\relax}
	\providecommand{\bibinfo}[2]{#2}
	\providecommand{\BIBentrySTDinterwordspacing}{\spaceskip=0pt\relax}
	\providecommand{\BIBentryALTinterwordstretchfactor}{4}
	\providecommand{\BIBentryALTinterwordspacing}{\spaceskip=\fontdimen2\font plus
		\BIBentryALTinterwordstretchfactor\fontdimen3\font minus
		\fontdimen4\font\relax}
	\providecommand{\BIBforeignlanguage}[2]{{%
			\expandafter\ifx\csname l@#1\endcsname\relax
			\typeout{** WARNING: IEEEtran.bst: No hyphenation pattern has been}%
			\typeout{** loaded for the language `#1'. Using the pattern for}%
			\typeout{** the default language instead.}%
			\else
			\language=\csname l@#1\endcsname
			\fi
			#2}}
	\providecommand{\BIBdecl}{\relax}
	\BIBdecl

\bibitem{ITU-T_G.975.1}
	ITU-T, ``Forward error correction for high bit-rate DWDM submarine systems,'' 2004. 
	[Online]. Available: www.itu.int/rec/T-REC-G.975.1

\bibitem{ITU-T_G.977}
	ITU-T, ``Characteristics of optically amplified optical fibre submarine cable systems,'' 2015. 
	[Online]. Available: www.itu.int/rec/T-REC-G.977

\bibitem{oda_2017}
	S.~Oda, M.~Miyabe, S.~Yoshida, T.~Katagiri, Y.~Aoki, T.~Hoshida, J.~C.~Rasmussen, M.~Birk, and K.~Tse, 
	``A learning living network with open {ROADMs},'' 
	\emph{{IEEE/OSA} J. Lightw. Technol.}, vol.~35, no.~8, pp.~1350--1356, Apr. 2017.

\bibitem{dong_2016}
	Z.~Dong, F.~N.~Khan, Q.~Sui, K.~Zhong, C.~Lu, and A.~P.~T.~Lau,
	``Optical performance monitoring: a review of current and future technologies,'' 
	\emph{{IEEE/OSA} J. Lightw. Technol.}, vol.~34, no.~2, pp.~525--543, Jan. 2016.

\bibitem{kilper_2004}
	D.~C.~Kilper, R.~Bach, D.~J.~Blumenthal, D.~Einstein, T.~Landolsi, L.~Ostar, M.~Preiss, and A.~E.~Wilner,
	``Optical performance monitoring,'' 
	\emph{{IEEE/OSA} J. Lightw. Technol.}, vol.~22, no.~1, pp.~294--304, Jan. 2004.

\bibitem{wu_2009}
	X.~Wu, J.~A.~Jargon, R.~A.~Skoog, L.~Paraschis, and A.~E.~Wilner, 
	``Applications of artificial neural networks in optical performance monitoring,'' 
	\emph{{IEEE/OSA} J. Lightw. Technol.}, vol.~27, no.~16, pp.~3580--3589, Aug. 2009.

\bibitem{fu_2005}
	B.~Fu, R.~Hui, and D.~L.~Richards,
	``Optical channel performance monitoring using coherent detection,'' 
	\emph{Opt. Fib. Commun. Conf. (OFC)}, Anaheim, CA, USA, Mar. 2005, Paper~OWJ2.

\bibitem{hauske_2009}
	F.~N.~Hauske, M.~Kuschnerov, B.~Spinnler, and B.~Lankl,
	``Optical performance monitoring in digital coherent receivers,'' 
	\emph{{IEEE/OSA} J. Lightw. Technol.}, vol.~27, no.~16, pp.~3623--3631, Aug. 2009.

\bibitem{adams_2006}
	R.~Adams, M.~Rochette, T.~T.~Ng, and B.~J.~Eggleton,
	``All-optical in-band {OSNR} monitoring at {40 Gb/s} using nonlinear optical loop mirror,'' 
	\emph{{IEEE} Photon. Technol. Lett.}, vol.~18, no.~3, pp.~469--471, Feb. 2006.

\bibitem{zhao_2014}
	D.~Zhao, L.~Xi, X.~Tang, W.~Zhang, Y.~Qiao, and X.~Zhang,
	``Periodic training sequence aided in-band {OSNR} monitoring in digital coherent receiver,'' 
	\emph{{IEEE} Photonics Journal}, vol.~6, no.~4, Aug. 2014.

\bibitem{bosco_2011}
	G.~Bosco, V.~Curri, A.~Carena, P.~Poggiolini, and F.~Forghieri,
	``On the performance of {Nyquist-WDM Terabit} superchannels based on {PM-BPSK, PM-QPSK, PM-8QAM or PM-16QAM} subcarriers,'' 
	\emph{{IEEE/OSA} J. Lightw. Technol.}, vol.~29, no.~1, pp.~53--61, Jan. 2011.

\bibitem{sinkin_2013}
	O.~V.~Sinkin, J.-X.~Cai, D.~G.~Foursa, G.~Mohs, and A.~N.~Pilipetskii, 
	``Impact of broadband four-wave mixing on system characterization,'' 
	\emph{{IEEE/OSA} Opt. Fib. Commun. Conf. and Exposition and The Natl. Fib. Opt. Eng. Conf. (OFC/NFOEC)}, 
	Anaheim, CA, USA, Mar. 2013, Paper~OM2B.4.2013, Paper OTh3G.3.

\bibitem{xiang_2018}
	Y.~Xiang, M.~Tang, Q.~Wu, H.~Zhou, B.~Yong, S.~Fu, D.~Liu,
	``A joint {OSNR} and nonlinear distortions estimation method for optical fiber transmission system,'' 
	\emph{{IEEE} Photonics Journal}, vol.~10, no.~5, Sep. 2018.

\bibitem{kashi_2018}
	A.~S.~Kashi, Q.~Zhuge, J.~C.~Cartledge, S.~Ali~Etemad, A.~Borowiec, D.~W.~Charlton, C.~Laperle, and M.~O'Sullivan, 
	``Nonlinear signal-to-noise ratio estimation in coherent optical fiber transmission systems using artificial neural networks,'' 
	\emph{{IEEE/OSA} J. Lightw. Technol.}, vol.~36, no.~23, pp.~5424--5431, Dec. 2018.

\bibitem{tani_2019}
	T.~Tanimura, K.~Tajima, S.~Yoshida, S.~Oda, and T.~Hoshida, 
	``Experimental demonstration of a coherent receiver that visualizes longitudinal signal power profile over multiple spans out of its incoming signal,'' 
	in \emph{Proc. Eur. Conf. Opt. Commun. (ECOC)}, Dublin, Ireland, Sep. 2019, Paper~PD.3.4.

\bibitem{bergano_1993}
	N.~S.~Bergano, F.~W.~Kerfoot, and C.~R.~Davidson,
	``Margin measurement in optical amplifier systems,'' 
	\emph{{IEEE} Photon. Technol. Lett.}, vol.~5, no.~3, pp.~304--306, Mar. 1993.

\bibitem{song_2002}
	L.~Song, M.-L.~Yu, and M.~S.~Shaffer,
	``10- and 40-{Gb/s} forward error correction devices for optical communications,'' 
	\emph{{IEEE} J. Solid-state circuits}, vol.~37, no.~11, pp.~1565--1573, Nov. 2002.

\bibitem{roberts_2009_jlt}
	K.~Roberts, M.~O'Sullivan, K.-T.~Wu, H.~Sun, A.~Awadalla, D.~J.~Krause, and C.~Laperle, 
	``Performance of dual-polarization {QPSK} for optical transport systems,'' 
	\emph{J. Lightw. Technol.}, vol.~27, no.~16, pp.~3546--3559, Aug.~2009.

\bibitem{chang_2010_cm}
	F.~Chang, K.~Onohara, and T.~Mizuochi, 
	``Forward error correction for {100 G} transport networks,'' 
	\emph{IEEE Commun. Mag.}, vol.~48, no.~3, pp.~S48--S55, Mar.~2010.

\bibitem{alvarado_2018_jlt}
	A.~Alvarado, T.~Fehenberger, B.~Chen, F.~M.~J.~Willems,
	``Achievable information rates for fiber optics: applications and computations,'' 
	\emph{{IEEE/OSA} J. Lightw. Technol.}, vol.~36, no.~2, pp.~424--439, Jan.~2018.

\bibitem{bocherer_2015}
	G.~B\"ocherer, F.~Steiner, and P.~Schulte, 	
	``Bandwidth efficient and
	rate-matched low-density parity-check coded modulation,'' 
	\emph{{IEEE} Trans. Commun.}, vol.~63, no.~12, pp.~4651--4665, Dec. 2015.

\bibitem{buchali_2016}
	F.~Buchali, F.~Steiner, G.~B\"ocherer, L.~Schmalen, P.~Schulte, and W.~Idler,
	``Rate adaptation and reach increase by probabilistically shaped {64-QAM}: an experimental demonstration,'' 
	\emph{{IEEE/OSA} J. Lightw. Technol.}, vol.~34, no.~7, pp.~1599--1609, Apr.~2016.

\bibitem{schmalen_2017}
	L.~Schmalen, A.~Alvarado, and R.~Rios-M\"uller,
	``Performance prediction of nonbinary forward error correction in optical transmission experiments,'' 
	\emph{{IEEE/OSA} J. Lightw. Technol.}, vol.~35, no.~4, pp.~1015--1027, Feb. 2017.

\bibitem{alvarado_2015}
	A.~Alvarado, E.~Agrell, D.~Lavery, R.~Maher, and P.~Bayvel, 
	``Replacing the soft {FEC} limit paradigm in the design of optical communication systems,''
	\emph{{IEEE/OSA} J. Lightw. Technol.}, vol.~33, no.~20, pp.~4338--4352, Oct. 2015.

\bibitem{cho_2017}
	J.~Cho, L.~Schmalen, and P.~J.~Winzer,
	``Normalized generalized mutual information as a forward error correction threshold for probabilistically shaped {QAM},''
	in \emph{Proc. Eur. Conf. Opt. Commun. (ECOC)}, Gothenburg, Sweden, Sep. 2017, Paper~M.2.D.2.

\bibitem{bocherer_2017_arxiv}
	G.~B\"ocherer,
	``Achievable rates for probabilistic shaping,''
	[Online]. Available: \url{arxiv.org/abs/1707.01134}

\bibitem{yoshida_2017_ptl}
	T.~Yoshida, M.~Karlsson, and E.~Agrell,
	``Performance metrics for systems with soft-decision {FEC} and probabilistic shaping,'' 
	\emph{{IEEE} Photon. Technol. Lett.}, vol.~29, no.~23, pp.~2111--2114, Dec. 2017.

\rev{
\bibitem{bocherer_2019_jlt}
	G.~B\"ocherer, P.~Schulte, and F.~Steiner, 
	``Probabilistic shaping and forward error correction for fiber-optic communication systems,'' 
	\emph{{IEEE/OSA} J. Lightw. Technol.}, vol.~37, no.~2, pp.~230--244, Jan.~2019.	
}	

\bibitem{yoshida_2019_ecoc}
	T.~Yoshida, M.~Mazur, J.~Schr\"oder, M.~Karlsson, E.~Agrell,
	``Performance monitoring for live systems with soft {FEC},'' 
	in \emph{Proc. Eur. Conf. Opt. Commun. (ECOC)}, Dublin, Ireland, Sep. 2019, Paper~W.3.D.5.

\bibitem{zhang_2019}
	S.~Zhang, F.~Yaman, E.~Mateo, I.~B.~Djordjevic, K.~Nakamura, T.~Inoue, and Y.~Inada,
	``On the performance metric and design of non-uniformly shaped constellation,'' 
	in \emph{Proc. Opt. Fib. Commun. Conf (OFC)}, San Diego, CA, USA, Mar. 2019, Paper~W1D.7.

\bibitem{yoshida_ecoc_2018}
	T.~Yoshida, M.~Karlsson, and E.~Agrell,
	``Technologies toward implementation of probabilistic constellation shaping,''
	in \emph{Proc. Eur. Conf. on Opt. Comm. (ECOC)}, Rome, Italy, Sep.~2018, Paper~Th.1.H.1.

\bibitem{yoshida_jlt_2019}
	T.~Yoshida, M.~Karlsson, and E.~Agrell,
	``Hierarchical distribution matching for probabilistically shaped coded modulation,''
	\emph{{IEEE/OSA} J. Lightw. Technol.}, vol.~37, no.~6, pp.~1579--1589, Mar.~2019.

\bibitem{amari_2019_arxiv}
	A.~Amari, S.~Goossens, Y.~C.~G\"ultekin, O.~Vassilieva, I.~Kim, T.~Ikeuchi, C.~Okonkwo, F.~M.~J.~Willems, and A.~Alvarado, 
	``Introducing enumerative sphere shaping for optical communication systems with short blocklengths,''
	[Online]. Available: \url{https://arxiv.org/pdf/1904.06601v3}

\bibitem{zehavi_1992_tcom}
	E.~Zehavi,
	``8-PSK trells codes for a Rayleigh channel,''
	\emph{{IEEE} Trans. Commun.}, vol.~40, no.~3, pp.~873--884, May~1992.

\bibitem{caire_1998_tit}
	G.~Caire, G.~Taricco, and E.~Biglieri,
	``Bit-interleaved coded modulation,''
	\emph{{IEEE} Trans. Inf. Theory}, vol.~44, no.~3, pp.~927--946, May~1998.

\bibitem{fabregas_2008}
	A.~Guill{\'e}n~i~F{\`a}bregas, A.~Martinez, and G.~Caire, 
	``Bit-interleaved coded modulation,''
	\emph{Found. Trends Commun. Inf. Theory}, vol.~5, nos.~1/2, pp.~1--153, 2008.

\bibitem{bicmbook}
	L.~Szczecinski and A.~Alvarado, 
	\emph{Bit-Interleaved Coded Modulation: Fundamentals, Analysis, and Design}. 
	Chichester, UK, John Wiley \& Sons, 2015.

\bibitem{400zr}
	OIF, ``400ZR,'' [Online]. Available: www.oiforum.com/technical-work/hot-topics/400zr-2/

\bibitem{openROADM}
	Open ROADM MSA, [Online]. Available: www.openroadm.org/home.html

\bibitem{lyub_2018}
	I.~Lyubomirsky, B.~Smith, J.~Riani, S.~Bhoja, 
	G.~Nicholl, M.~Nowell, F.~Villarruel, 
	``{400G FEC} and framing for 80 km,'' [Online]. 
	Available: www.grouper.ieee.org/groups/802/3/B10K/public/18\_09/lyubomirsky\_\\
	b10k\_01a\_0918.pdf

\bibitem{cho_2019_jlt}
	J.~Cho and P.~J.~Winzer,
	``Probabilistic constellation shaping for optical fiber communications,''
	\emph{{IEEE/OSA} J. Lightw. Technol.}, vol.~37, no.~6, pp.~1590--1607, Mar.~2019.

\bibitem{yoshida_2019_arxiv}
	T.~Yoshida, A.~Alvarado, M.~Karlsson, and E.~Agrell,
	``Post-{FEC} {BER} prediction for bit-interleaved coded modulation with probabilistic shaping,'' 
	[Online]. Available: \url{arxiv.org/abs/1911.01585}

\rev{
\bibitem{yoshida_2020_ofc}
	T.~Yoshida, M.~Mazur, J.~Schr\"oder, M.~Karlsson, and E.~Agrell,
	``On the performance under hard and soft bitwise mismatched-decoding,''
	in \emph{Proc. Opt. Fib. Commun. Conf. (OFC)}, San Diego, CA, Mar. 2020, Paper.~Th1I.5.
}

\rev{
\bibitem{kaplan_1993_aeu}
	G.~Kaplan and S.~Shamai, 
	``Information rates and error exponents of compound channels with application to antipodal signaling in a fading environment,''
	\emph{Archive for Electronics and Transmission Technique (Archiv f\"ur Elektronik und \"Ubertragungstechnik, {AE\"U})}, vol.~47, no.~4, pp.~228--239, 1993.
}


\bibitem{yoshida_2016}
	T.~Yoshida, K.~Matsuda, K.~Kojima, H.~Miura, K.~Dohi, M.~Pajovic, T.~Koike-Akino, D.~S.~Millar, K.~Parsons, and T.~Sugihara,
	``Hardware-efficient precise and flexible soft-demapping for multi-dimensional complementary {APSK} signals,'' 
	in \emph{Proc. Eur. Conf. Opt. Commun. (ECOC)}, D\"usseldorf, Germany, Sep. 2016, Paper~Th.2.P2.SC3.27.

\bibitem{schulte_2016}
	P.~Schulte and G.~B\"ocherer, ``Constant composition distribution matching,''
	\emph{{IEEE} Trans. Inf. Theory}, vol.~62, no.~1, pp.~430--434, Jan. 2016.

\bibitem{sugihara_2013}
	K.~Sugihara, Y.~Miyata, T.~Sugihara, K.~Kubo, H.~Yoshida, W.~Matsumoto, and T.~Mizuochi,
	``A spatially-coupled type {LDPC} code with an {NCG} of 12 {dB} for optical transmission beyond {100 Gb/s},'' 
	in \emph{Proc. Opt. Fib. Commun. Conf. (OFC)}, Anaheim, CA, USA, Mar. 2013, Paper~OM2B.4.

\bibitem{mazur_2019}
	M.~Mazur, J.~Schr\"oder, A.~Lorences-Riesgo, T.~Yoshida, M.~Karlsson, and P.~A.~Andrekson,
	``Overhead-optimization of pilot-based digital signal processing for flexible high spectral efficiency transmission,'' 
	\emph{OSA Opt. Express}, vol.~27, no.~17, pp.~24654--24669, Aug. 2019.

\bibitem{yoshida_2018_ofc_spg}
	T.~Yoshida, M.~Karlsson, and E.~Agrell,
	``Low-complexity variable-length output distribution matching with periodical distribution uniformalization,''
	in \emph{Proc. Opt. Fib. Commun. Conf. (OFC)}, San Diego, CA, Mar. 2018, \rev{Paper}~M4E.2.

\end{thebibliography}

\end{document}